\documentclass{article}
\usepackage{graphicx}  
\usepackage{amsmath}   
\usepackage[compress]{cite}
\usepackage{amssymb}   
\usepackage{bm} 
\usepackage{dcolumn}
\usepackage{color}
\usepackage{mathrsfs}
\usepackage{amsfonts}
\usepackage{varioref}
\usepackage{textcomp}
\RequirePackage[colorlinks,citecolor=blue,urlcolor=magenta,linkcolor=blue]{hyperref}
\allowdisplaybreaks
\labelformat{section}{Section #1} 
\labelformat{subsection}{Section #1} 
\labelformat{subsubsection}{Section #1}
\labelformat{subsubsubsection}{Section #1}
\labelformat{equation}{Eq.~(#1)} 
\labelformat{figure}{Fig.~#1} 
\labelformat{subfigure}{Fig.~\thefigure#1} 
\labelformat{table}{Tab.~#1} 
\labelformat{appendix}{Appendix #1}
\title{\Large \bf Information content and minimum-length metric: A drop of light}
\author{ Alessandro Pesci \footnote{pesci@bo.infn.it}
\\
{\small{INFN Bologna, Via Irnerio 46, I-40126 Bologna, Italy}}
}
\date{ }  

\begin{document}

\maketitle

\begin{abstract}
In the vast amount of results linking gravity
with thermodynamics, statistics, information,
a path is described which tries to explore this connection
from the point of view of (non)locality
of the gravitational field.
First the emphasis is put on
that well-known thermodynamic results
related to null hypersurfaces (i.e. to lightsheets and to generalized covariant
entropy bound) can be interpreted as implying
an irreducible intrinsic nonlocality of gravity.
This nonlocality even if possibly concealed at ordinary scales
(depending on which matter is source of the gravitational field,
and which matter we use to probe the latter)
unavoidably shows up at the smallest scales, read the Planck length $l_p$,
whichever are the circumstances we are considering.

Some consequences are then explored of 
this nonlocality 
when embodied
in the
fabric itself of spacetime 
by endowing the latter
with a minimum length $L$,
in particular the well-known and intriguing fact 
that this brings
to get the field equations, and all of gravity with it, 
as a statistical-mechanical result. 
This is done here
probing the neighborhood of a would-be (in ordinary spacetime)
generic event 
through lightsheets 
(instead of spacelike or timelike geodesic congruences
as in other accounts) from it.
The tools for these derivations
are nonlocal quantities,
and among them
the minimum-length Ricci scalar stands out 
both for providing
micro degrees of freedom for gravity in the statistical account
and for the fact that intriguingly
the ordinary, or `classical', Ricci scalar can not
be recovered from it in the $L\to 0$ limit.  

Emphasis is put on that classical gravity 
is generically obtained this way for $\hbar\ne 0$, 
but not in the $\hbar\to 0$ limit
(the statistically derived field equations become singular in this limit),
adding to previous results in this sense. 
This hints to that
the geometric description of 
gravity we are used to
is intrinsically quantum,
as it requires $\hbar \ne 0$,
on top of being of statistical-mechanical origin. 
One would expect that this inherently non-classical nature 
might show up through nonlocality
also at scales much larger than $l_p$ if special suitable
circumstances are considered.
\end{abstract}


\tableofcontents


$ $
\section{Preamble}

Gravity and thermodynamics:
It is definitely not an easy task to write on these matters,
since by direct experience we all know
these were topics of utmost significance for 
Prof. Thanu Padmanabhan (in memory of which I am writing
and which I would like to call here affectionately
Paddy), and we lost him.

My first acquaintance with him has been 
through consideration of some of his deeply inspiring works
\cite{PadI, PadJ, PadK} about gravity as an emergent phenomenon,
and in occasion
of a paper \cite{PesG} it had occurred to me to post
about connections between gravity and thermodynamics,
with a perspective in which 
--to my most complete surprise, but not such an uncommon thing
as I realized afterwards for it happened to many others-- 
he contacted me to say he saw in it something interesting.
Since then, there has been a certain number 
of opportunities for me to discuss with him
new results on these topics,
almost always coming these from his side. 
What I would focus on in this paper however,
is some (few) results from my side
and the line of research they seem to suggest.
I would dare to do this way in part because
these are after all the things I can talk about;
in part, and more to the point, 
also because it was Paddy himself
to find (quite not convincingly)
some value in them;
I would feel thus  as having
kind of green light by him on this in a paper in his memory.

What the paper deals with is
gravity and thermodynamics in the broadest sense. 
There is absolutely no intent (nor capability from my side)
to give an exhaustive description of such an
immense field of research.
%
What here reported is, on the contrary, a specific, particular
path through the field, reflecting my personal feelings and interests
often entangled with some of Paddy's last years achievements.
There will then be plenty of important results in the field 
not mentioned here (including many results from Paddy himself). 
They are of course really relevant, and this is only 
unavoidably due to the specificity of the path (and my lack). 
I apologize for this. 

The plan of the paper is as follows. 
Section 2 deals with spacetime and field equations
with an emphasis on elements 
which bring thermodynamics into the scene,
in spite of never being called upon 
in the conventional derivation of field equations. 
As Paddy strongly emphasized
(see e.g. \cite{PadI, PadJ, PadK} and \cite{PadN}, 
and references therein), 
this gives to gravity a thermodynamic flavor,
hinting to some kind of emergent nature for it,
and suggests there should be a way to look 
at field equations as what results from kind of
max entropy procedure. 
No statistical mechanics can be present
at this stage,
since there is complete ignorance of the
fundamental microscopic degrees of freedom (or dofs)
from which dynamics
gravity would be suspected to emerge.

Section 3 deals with the discovery 
\cite{Pad01, Pad02} that
consistently endowing spacetime with a lower limit length
is enough to provide microscopic (or mesoscopic)
dofs.
And how using these,
a statistical-mechanical derivation of field equations
becomes possible.
Some intriguing aspects of the expression
of the Ricci scalar in the minumum-length metric
are discussed,
in particular the fact that classical spacetime
can not be recovered
in the limit in which this minimal length
is vanishingly small.

\section{Spacetime and thermodynamics}

\subsection{Thermodynamics at the centre: Jacobson's result}

The insight of Bekenstein 
\cite{BekensteinHorizonA, BekensteinHorizonB, BekensteinHorizonC} 
of ascribing  to black hole
horizon an entropy,
as well as 
the key result by Hawking \cite{HawkingA, HawkingB}
of finding 
(combining quantum mechanics and general relativity)
a temperature for it,
brought thermodynamics in gravity at a fundamental 
or constitutive level.
It gave the laws of black hole mechanics \cite{Bardeen}
turned into laws of thermodynamics.
This was, and is, a mind-boggling feature,
for Einstein's field equations are purely geometric 
in their solutions,
and one does not know how it is that an intrinsically geometric
concept like that of a horizon of a black hole,
can have an inherently statistical meaning
with an associated entropy $S_H = {A_H}/4$
where $A_H$ is the area of the horizon in Planck units
(we use these units if not explicitly stated otherwise).

Later works further strengthened the case
and brought all this to a context more general
than just black holes.
\begin{figure}[h!]
  \includegraphics[width = 8 cm]{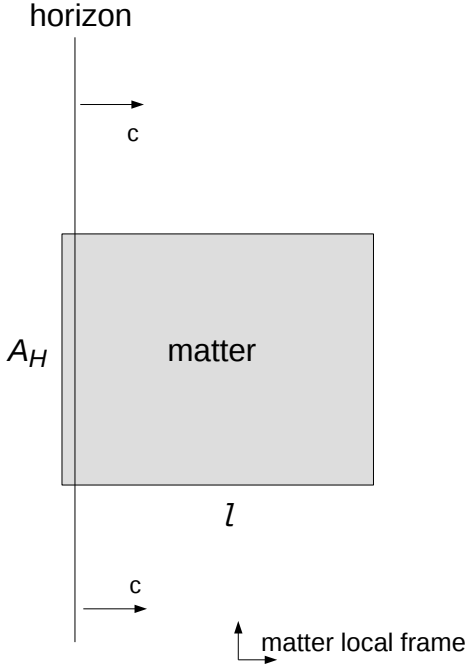}
  \caption{Crossing of the (Rindler) horizon by a lump of matter as described
           in matter local frame.}
  \label{fig}
\end{figure}
In particular, 
Jacobson \cite{Jacobson}
considered a patch of Rindler horizon 
\cite{Rindler}
around an arbitrary event $P$, 
with no expansion and shear at first order
at $P$
(`local Rindler horizon').
For it,
he went to show that
if we assume that any patch of the perceived horizon
can be endowed, from the mere fact of having an area $A_H$,
by an entropy $S_H = {A_H}/4$, 
then Einstein's equations themselves are equivalent
to thermodynamic Clausius relation as applied to the horizon.
Indeed, Jacobson's argument can be given the form

\begin{eqnarray}\label{nullq44.3}
\delta Q_H  &=&  T_{ab} k^a k^b \, {A_H} \, \, l 
\, = \, T_H \delta S_H \, = \,
\frac{\kappa}{2 \pi} \, \frac{1}{4} \, \delta {A_H}
\nonumber \\ 
&=&
\frac{\kappa}{2 \pi} \,  \frac{1}{4} \, \frac{1}{\kappa} \,
R_{ab} k^a k^b \, {A_H} \, \, l 
\, = \,
\frac{1}{8 \pi} \, R_{ab} k^a k^b \, {A_H} \, \, l
\, = \,
\frac{1}{8 \pi} \, R_{ab} k^a k^b \, V,
\end{eqnarray}
which stems from 
taking $T_H = \kappa/(2 \pi)$ 
\cite{Unruh}
as  horizon's temperature  
with $\kappa$ the acceleration
of the observer (third equality), 
and from use of Raychaudhuri's equation
as applied to the local Rindler horizon (fourth equality).
Here the symbol $\delta$ 
denotes variations of horizon's quantities
resulting from the swallowing of matter.
$\delta Q_H$ is the total energy supplied to the horizon 
by the swallowing  of an element of matter of small area $A_H$ and 
proper thickness $l$, 
as measured in matter rest frame,
then of proper volume $V = {A_H} \, l$ (Fig. \ref{fig});
$T_{ab}$ is the stress-energy tensor, $R_{ab}$ the Ricci tensor 
and $k^a = (d/d\lambda)^a$ are the generators
of the Rindler horizon with the affine parameter $\lambda$ taken
to be distance or time in the local frame of matter.

\subsection{Holography: Lightsheets and the (generalized) covariant 
entropy bound (GCE bound)}

Consideration of the fact that any lump of matter/energy dumped 
to a black hole brings to an increase of the area
of the horizon, and of the evidence that any localized amount of 
entropy/information has with it a finite energy (as
implied e.g. by the Bekenstein bound \cite{BekensteinSpecificEntropy}), 
brings generically to expect that 
the entropy of a black hole is actually the
maximum entropy one can have inside a spherical 2-surface with area equal 
to black hole’ horizon \cite{'t Hooft, Susskind}.
This has been the basis of the conjecture 
that all the physics in the bulk of a $D$-dim region, can
be encoded in dofs living on the $(D-1)$-dim boundary 
(dimensional reduction or holographic principle)
\cite{'t Hooft, Susskind}, 
which has found a concrete implementation
in string theory in the AdS/CFT correspondence \cite{Maldacena}
and in a number of similar examples of dualities. 

The generalized covariant entropy bound \cite{BoussoCov, FMW}
\begin{eqnarray}\label{genBousso}
S(L) \, \le \, \frac{{A_B} - {A_{B'}}}{4},
\end{eqnarray}
can be considered as the most general formulation of the holographic principle 
in Einstein spacetime.
Here,
$B$ is a generic 
spacelike 2-dim surface with area $A_B$, from which
light rays emanate orthogonally 
in nonexpanding directions.
The rays are followed until they reach
another 2-dim spacelike surface $B'$ (area $A_{B'}$) 
orthogonal to them
where they are terminated before focal points are reached
(the (non-generalized) covariant entropy bound is the particular case
when they are allowed to reach the focal points and $A_{B'} = 0$).
$L$ is
the null 3-dim submanifold swept by the light rays,
and $S(L)$ is matter system' entropy on $L$. 
 
Like Einstein's field equations, 
the bound relates geometry to matter dofs,
but, intriguingly, the latter appear now through 
their possibly full information content instead of energy.
One has then good reasons to
suspect that a complete account
of the bound, 
assuming it to hold true down at the fundamental level,
must wait until one will have
a complete theory of gravity and matter.
On the positive side,
any attempts to prove or check, or possibly modify, the bound
at an assigned level of description
might tell something about the nature of gravity, geometry 
and matter at a deeper level.  

Even if it exhibits geometric quantities,
the expression to the right in bound (\ref{genBousso})
is actually an entropy
(as it ought to be, from the meaning of the left-hand side).
The result \cite{Unruh} indeed, 
which
the same as for black-hole horizons  
assigns a temperature
to any local Rindler horizon,
allows to
extend the notion of temperature 
to arbitrary null hypersurfaces.
A surface element of $B$ or $B'$ with area $dA$
can act as a spatial section
of a local Rindler horizon of an orthogonally accelerating observer,
and this horizon
can be assigned a temperature (dependent on the acceleration),
and an entropy $dS_H  = \frac{dA}{4}$ independent of acceleration 
(explicitly inserting Newton's constant $G$,
$dS_H  = \frac{dA}{4 G}$ in general relativity 
which is where bound (\ref{genBousso}) lives, 
and $dS_H  = \frac{dA}{4 G_{\rm eff}}$ 
(Wald entropy \cite{Wald, JKM, IyerWald})
in more general theories with $G_{\rm eff}$ an effective coupling constant
replacing $G$ and in general varying with the point 
(e.g. through dependence on curvature) 
\cite{BrusteinGorbonosHadad}).
We can regard $L$ in a neighborhood of $B$ ($B'$)
as a horizon with spatial section $B$ ($B'$).
This horizon can be approximated
at each $p\in B$ ($B'$)
by a local Rindler horizon  
with generators $k^a$ along the light ray at $p$.
These local horizons can be chosen to have all a same temperature,   
and all possess 
entropies $dA/4$. 

This shows that the fact that spacetime does have null hypersurfaces
brings with it that to the latter 
is associated heat density, and entropy.
Right this observation, we know, forms the basis
of Paddy's emergent gravity
paradigm \cite{PadI, PadJ}.   
In it, extending the scope of Boltzmann's intuition to include spacetime,
the idea is that if a null hypersurface can be hot 
it must have microscopic dofs.
This suggests that spacetime as we know it might then arise
from these latter
with field equations being analogous
to the equations of fluid mechanics. 

Proofs of the bound (\ref{genBousso}) 
do exist \cite{FMW, BFM, StromingerThompson, PesB}
within the context of  classical general relativity
and for matter in the hydrodynamic regime,
this meaning circumstances with a classical spacetime
and in which matter quantum dofs 
can be described in terms of continuous variables,
like energy density $\rho$, entropy density $s$ with a local entropy
current, pressure $p$, temperature $T$, ..., at any point of it .

Bound (\ref{genBousso}) however is found to be true also
for matter considered beyond any restriction 
to the hydrodynamic approximation.
In \cite{quantumBoussoProof} the bound 
is proven indeed in the far more general context
of local quantum field theories,
even if restricting to the case of matter
consisting of free fields,
with the entropy on the lightsheet calculated as vacuum subtracted
von Neumann entropy for these fields,
and in the limit in which the gravitational backreaction is weak.
This proof does not rely on the null energy condition,
condition the latter
which one might think to generically accompany any description
of matter  
in which entropy and energy are microscopically related
as in the case of hydrodynamic approximation.
It accommodates then also circumstances in which the null energy
condition is locally explicitly violated as for example in the case of Hawking 
radiation at the horizon,
case in which possible violations of bound (\ref{genBousso})
have been pointed out \cite{Lowe}.
This has shown explicitly that the bound (\ref{genBousso})
holds true as it is also in these circumstances, 
apparently without strictly a need for the modifications of it 
envisaged in \cite{StromingerThompson}.
The proof \cite{quantumBoussoProof} is quite in the same vein
as the local-quantum-field-theory proof \cite{Wall1, Wall2} 
of the closely related generalized second law
(even if with the relevant entropies being however meant differently
in the two contexts: entropy on the part of the horizon 
to the future of any given spatial section in case of \cite{Wall1, Wall2},
and entropy on the lightsheet between two spatial sections
in case of \cite{quantumBoussoProof}).  

Our aim here is not so much to focus
on further checks of thermodynamic bound (\ref{genBousso}) or on its proofs,
but, rather, on what we can infer about gravity
when taking the bound as true.
This would be in the spirit of the general efforts 
aiming to understand what is telling us holography 
about the nature of spacetime.
While doing this
we choose moreover to stay in the hydrodynamic approximation.
Clearly,
one would push the description, 
and thus likely the understanding,
to a level as deep and general as possible,
thus e.g. to quantum field theory description  
or to string theory gauge/gravity duality mentioned above.
Thanks to the fact, however,  
that the bound keeps a same formulation
both in quantum field theory and in the hydrodynamic regime,
there might be a point 
in trying to infer something 
about gravity also simply staying in the latter, 
which presents features perhaps 
more easy to manage. 
A question we might want to try to address for example
is what kind of mechanism or principle might be at work,
if any,
preventing bound (\ref{genBousso}) to be violated
while we keep staying in the hydrodynamic regime.

\subsection{Attaining the bound}

The proofs in \cite{FMW, BFM, StromingerThompson} are in terms of 
some reasonable conditions, obeyed by matter entropy,
which turn out to be sufficient for the validity of
the bound.
These conditions are considered reasonable with reference to 
actual, physical hydrodynamic systems.
Proof \cite{PesB} goes very much along the same line of effort
pushing it up to the point to try to identify
kind of extreme condition, 
the most challenging one, 
that the hydrodynamic matter 
has to satisfy,
violating which 
(though still safely remaining within the hydrodynamic regime)
would mean to violate the bound.

A convenient way to express this condition
is in terms of a systems consisting of a layer
of matter, i.e. occupying a volume in which two dimensions
are much larger than the third one, the thickness $l$.
In case of a homogeneous system 
made of ultrarelativistic particles or of photons
the condition takes the very simple form \cite{PesB}

\begin{eqnarray}\label{From.12}
l \, \ge \, \frac{1}{\pi T}
\, \, \, \Big[ = \frac{\hbar c}{\pi k T}\Big],
\end{eqnarray}
with $T$ the temperature of the system.
Here (and hereafter) between square brackets we 
explicitly reinsert
all the constants involved, namely, in present case, 
(reduced) Planck's, and Boltzmann's constants $\hbar$ and $k$ 
and the speed of light $c$. 
This to explicitly remark here
that this result comes about with gravity playing no role,
as indicated by the absence of Newton's constant $G$.  

In the general case
with constituent particles of generic nature,
one gets 
in an analogous manner
\cite{PesC}

\begin{eqnarray}\label{Onthe.6}
l \, \, \ge \, \, 
l^*({\rm medium}, {\rm state}) 
\, \, \equiv \, \, 
\frac{1}{\pi} \, \frac{s}{\rho + p}
\, \, = \, \,
\frac{1}{\pi T} \, \Big(1 - \frac{\mu n}{\rho +p}\Big),
\end{eqnarray}
with last expression applying in case the system is homogeneous.
Here 
$s$ and $n$ are respectively
local entropy and number densities,
and 
$\rho$ is local energy density,
including the rest mass of constituent particles;
$p$ is pressure, 
and $\mu$ the chemical potential
which also consistently includes the rest mass energy
of the particles
and can thus be expected to be $\mu \ge 0$.
The equality follows from Gibbs-Duhem relation
$\rho = T s - p + \mu n$. 
The length $l^*$ is 
characteristic of the medium which the system is made of
and depends moreover
on the thermodynamic state corresponding to the assigned values
of the thermodynamic potentials;
for a photon gas at temperature $T$, 
$\mu = 0$ and $l^* = \frac{1}{\pi T}$ in agreement with (\ref{From.12}).  
$l^*$ sets the limit below which the generalized covariant entropy
bound might ideally be violated at the assigned thermodynamic
circumstances, 
still remaining safely within the hydrodynamic regime.   
More precisely,
if we consider
a spacelike two surface so chosen to have
orthogonal rays with vanishing expansion  
and we put on it
a layer of matter with a thickness $l$ smaller
than the $l^*$ characteristic of that matter at the assigned 
values of thermodynamic potentials,
then if we terminate the light-sheet at the exit of the layer
we would obtain a violation of bound (\ref{genBousso}).

What actually happens \cite{PesC} is that
making an inventory of the possible thermodynamic systems
in a wide range of values of thermodynamic potentials,
$l^*$ turns out invariably to be by far smaller 
than the `size' $\lambda$ of constituent particles themselves, 
taken this size to
be their intrinsic quantum indeterminacy at the thermodynamic
conditions.   
Since the thickness $l$ of the layer by definition 
can not be smaller than $\lambda$,
what we generically have is 
$l \ge \lambda \gg l^*$, and the bound (\ref{genBousso}) is safe. 

Looking at (\ref{Onthe.6}),
if we want to challenge the bound we have to search for systems
which have a high $s$ gain for given $\rho$,
or high $s/\rho$ ratio, 
in other words we are (of course) interested in the most entropic systems, 
and this means that
we can not help thinking of a thermal photon gas.
Indeed, 
if we consider a photon gas at thermal equilibrium at temperature $T$
and assume that $\delta p = \epsilon/2$,
with $\epsilon = 2.82 \, T$ the peak energy 
of photon distribution \cite{KittelKroemer},
captures the intrinsic indeterminacy in the momentum of the photons,
then
from 
$\delta x \, \delta p \ge 1/2$ with $\delta x$ the indeterminacy in
photon position, 
we get 
$\lambda = \delta x \ge 
1/\epsilon = \frac{1}{2.82 \, T} 
\approx \frac{1}{\pi T} = l^*$. 
Thus, ideally, a very thin layer of this gas,
as thin as possible compatibly with the intrinsic spatial indeterminacy
of the photons at temperature $T$, almost does the job, 
it attains the bound (\ref{genBousso}) \cite{PesC}.
Layers thinner than this can be conceived,
but due to uncertainty relations 
they necessarily
would be accompanied by a temperature larger than the given $T$.
There is (almost, as we will consider later) 
no lower limit 
to the size 
at which we can apply
our analysis, but invariably for that thinner system we just attain
the bound as just showed, we can not beat it. 
The reason why 
we can not violate the bound,
is that this would mean to violate the indeterminacy relations;
there are no constraints instead coming from our choice
to remain within the hydrodynamic regime. 
Quantum indeterminacy leads to (\ref{From.12}) and (\ref{Onthe.6});
once this is given, 
then consideration of gravity (Einstein's equations) brings 
to the bound (\ref{genBousso}).

\subsection{Hod's bound to relaxation times}

Before we proceed
let us pause to mention some ramifications of this result
with connections to apparently far-apart topics,
a fact which might deserve further investigation.
One of these is the relation with the relaxation times
of thermodynamic systems, namely the characteristic times
for a system put slightly off equilibrium to return 
to the equilibrium state. 
For any system we can think of its relaxation time $\tau$ 
as $\tau \ge d/v_s$ where $d$ is the size of the system 
and $v_s$ is the speed of sound in it, 
with the limit
being in principle attainable 
in the most favorable geometrical circumstances. 
Coming back to the just considered idealized system consisting of
an as-thin-as-possible layer of a photon gas at temperature $T$,
namely with thickness $l = \lambda = \frac{1}{\pi T} = l^*$,
if we take as $v_s$ the vacuum speed of light, $v_s = 1$, 
we get $\tau = \frac{1}{\pi T}$ \cite{PesD}.    

Now, from thermodynamics used
within quantum information theory,
a universal lower limit $\tau_{min}$ had been envisaged before precisely
of the form
$\tau_{min} = \frac{1}{\pi T}$ 
\big[$ = \frac{\hbar}{\pi k T}$\big] \cite{Hod1}.
We can see (as first noticed by Hod \cite{Hod2})
that the bound associated to this limit 
and bound (\ref{From.12}) do coincide
with length replacing time,
this implying that the highly idealized layer we chose
in order to attain the generalized entropy bound (\ref{genBousso})
actually attains also the universal bound to relaxation times \cite{PesD}.
Hod's bound is --from its derivation or looking directly
at the physical constants involved in the result--
independent from gravity, the same as we emphasized for (\ref{From.12}). 
Still, quite intriguingly,
black holes are theoretically found
(from consideration of quasi-normal modes 
of their free oscillations when perturbed)
to give relaxation times of the order of magnitude of Hod's limit
and conform to it \cite{Hod1}.
This appears now quite confidently {\it experimentally} confirmed
from the analysis \cite{Carullo}
of the data of black hole mergers
taken by LIGO-Virgo Collaboration, 
as it can be seen starting already from
the first \cite{LIGOVirgo} detected event GW150914.   
The extremal black holes
are expected to actually saturate the bound \cite{Hod1}.

\subsection{KSS bound to viscosity/entropy ratio}

Another topic which the bounds (\ref{From.12}) and (\ref{Onthe.6})
turn out to be 
related to 
is the so called KSS (Kovtun, Son, Starinets) bound
to viscosity/entropy \cite{KSS1, KSS2},
put forward in the context of AdS/CFT.
In it,
on the basis of the result that thermal theories 
with gravity dual invariably exhibit 
a same ratio $\eta/s = \frac{1}{4 \pi}$,
where $\eta$ is viscosity and $s$ entropy density,
and that for thermal theories which are instead generic 
this ratio is in general much larger, the conjecture is made
that 
$\eta/s \ge \frac{1}{4 \pi}$ 
\big[$\eta/s \ge \frac{\hbar}{4 \pi k}$\big]
for any system which can be described through a consistent
relativistic quantum field theory.  
Many counterexamples have been found for this bound
when including higher derivative corrections,
not only in the sense of lowering the factor 1/4,
but also showing that the ratio $\eta/s$ can be made arbitrarily
small, typically this involving systems with arbitrarily large
number of species \cite{Cremonini}
(but not being restricted to this;
a very recent account of possible violations of the bound in general settings
can be found in \cite{Meert}).
%
However, at least in some of these cases consistency issues 
can then be raised,   
and it seems anyway 
that, at least when chemical potential $\mu = 0$,
$\eta/s$ cannot be vanishingly small 
but must be bounded by some factor
$\ne 0$ if not $\frac{1}{4 \pi}$ \cite{Lawrence}.  

As for our purposes here, 
we emphasize that the KSS bound can clearly be reframed
as an entropy bound \cite{BekensteinViscosity}.
It is then by now no surprise that it can be related
to bounds (\ref{From.12}, \ref{Onthe.6}).  
What happens is that our as-thin-as-possible layer above
does again the job: 
in a back-on-the-envelope calculation
it attains the KSS bound,
with constant = $\frac{1}{4 \pi}$; 
for this, we have however to think
of strongly coupled massless particles, e.g. gluons, 
as replacing photons \cite{PesF}.   
Indeed
from $\eta \approx \frac{1}{3} t_c \rho$ \cite{Misner}
for radiation, with $t_c$ the collision time,
we get 

\begin{eqnarray}\label{Asemiclassical.5}
\frac{\eta}{s} \, \approx \, \frac{1}{3} \, t_c \, \frac{\rho}{s}
\, = \, \frac{1}{4} \, t_c \,  \frac{\rho + p}{s}
\, = \, \frac{1}{4 \pi} \, \frac{t_c}{l^*}
\, = \, \frac{1}{4 \pi} \, \frac{l_c}{l^*},
\end{eqnarray}
with $l_c$ the collision length \cite{PesE}.
Here $l_c \ge \lambda$, for in a volume $\lambda^3$ we have one
radiation quantum.
We can then leave $l_c$ to ideally become as small as $\lambda$,
but this requires the radiation is strongly 
coupled.\footnote{
Here, and in the quoted expression $\eta \approx \frac{1}{3} t_c \rho$
\cite{Misner} for viscosity,
it might seem at first puzzling that $\eta$ results higher for larger $t_c$,
thus for weaker coupling.
This however can be fairly understood thinking to that
$\eta$ is a measure of the rate of momentum diffusion,
and clearly the momentum transfer is easier the longer
the particle can travel between collisions (cf. e.g. \cite{SonStarinets}).}  
In fact,
starting from the Thomson cross-section
$
\sigma_{\rm T} =
\frac{8}{3} \pi \frac{\alpha^2}{m^2} =
\frac{2}{3\pi} \alpha^2 \lambda_{\rm Compton}^2 =
O(1) \, \alpha^2 \lambda_{\rm Compton}^2
$ 
of a particle with Compton wavelength $\lambda_{\rm Compton}$,
and assuming in analogy with it
the cross-section $\sigma$ of our radiation quanta to be
$
\sigma = O(1) \, g^2 \lambda^2,
$
we have
$
l_c = \frac{1}{\sigma n} = O(1) \, \frac{\lambda}{g^2}.
$
And this shows that to get on average a collision in
a length $\lambda$ we must have $g^2 = O(1)$,  \cite{PesF}.  
A different argument leading to the KSS limit
still using, among other things, 
the $l^*$ concept is presented in \cite{Hod3, Hod4}. 

In spite of having been originally derived
for quantum field theories dual to bulk stringy gravitational theories, 
the KSS bound has no intrinsic reference to gravity
as can be envisaged from its formulation, not containing $G$.
It is then similar in this respect to Hod's bound on relaxation times
and it also exhibits, 
as it happens for the latter, 
the intriguing feature that 
even if the bound knows at a fundamental level nothing of gravity,
black holes actually attain it.
It is indeed a (much earlier than AdS/CFT) result 
of membrane paradigm 
the fact that \cite{Damour1, Damour2}
the (surface) viscosity of Schwarzschild black hole
is
$\eta = \frac{1}{16 \pi} \, \big[ = \frac{1}{16 \pi} \, \frac{c^3}{G}\big]$
which, 
being the (surface) entropy density 
$s = \frac{1}{4} \, \big[ = \frac{1}{4} \, \frac{k c^3}{\hbar G}\big]$,
gives
$\frac{\eta}{s} = \frac{1}{4 \pi}$.

\subsection{Bekenstein's bound to entropy}

A third topic we wish to consider here in connection
with bounds (\ref{From.12}) and (\ref{Onthe.6}) above is
the Bekenstein bound \cite{BekensteinSpecificEntropy}.
We already generically used of it while introducing the spherical
entropy bound.
It reads \cite{BekensteinSpecificEntropy}

\begin{eqnarray}\label{BekensteinBound}
S \, \le \, 2 \pi \, E R 
\, \, \, \Big[ = 2 \pi \, E R \, \frac{k}{\hbar c}\Big]
\end{eqnarray}
with $S$  and $E$ the entropy and the rest energy of the system 
and $R$ the radius
of the smallest circumscribing sphere,
assuming that if spacetime nonnegligibly deviates 
from Minkowski, consideration is restricted to spacetimes which 
enjoy symmetries which allow for
these quantities (or at least the combination which appears
in the bound \cite{Casini})
to be meaningfully defined. 

Within the hydrodynamic regime and
for negligible selfgravity 
we can express this bound in terms of densities,
as

\begin{eqnarray}\label{BekensteinDensities}
s \, \le \, 2 \pi \, \rho R. 
\end{eqnarray}
We can recognize this has quite a resemblance to bound
(\ref{Onthe.6}) on thermodynamic densities
rewritten as 

\begin{eqnarray}\label{Onthe.5}
s \, \le \, \pi \, (\rho + p) \, l, 
\end{eqnarray}
except that the bound (\ref{Onthe.5}) 
is in general much stronger than (\ref{BekensteinDensities})
since we can take $l \ll R$ \cite{PesE}.
Given a material medium,
e.g. the quite miraculous photon gas considered above,
we can imagine to make balls out of it
of decreasing radius while keeping intact the values of
thermodynamic potentials.
Each ball can be considered as a thermodynamic system by its own,
the smaller the radius
the nearer coming it to attain the Bekenstein bound (\ref{BekensteinDensities}).
We can consider a radius $\bar r$ as small as needed to just
attain the Bekenstein bound. 
It is
$
\bar r = \frac{1}{2 \pi} \, \frac{s}{\rho}.
$
As easily verified,
it relates to the minimum allowed $l$ 
($l = l^* = \frac{1}{\pi T}$)
for the prescribed
values of the thermodynamic potentials 
of the thermal photon gas 
as 
$
l^* = \frac{3}{2} \, {\bar r}
$.
This quite nicely fits
with the limiting quantum-mechanical condition (\ref{From.12})  
for the size.  
Indeed,
a radius $r$ say with $l^* = 2 \, r$ would be too small
to comply with (\ref{From.12}), for any intercept with 
the ball would have length $l \le l^*$;
a radius with $l^* = r$ would be a little too large than strictly needed:
most of intercepts would have $l > l^*$, and a smaller ball
could be made still on average complying with (\ref{From.12}). 
It is then fair to say that
the thermal photon gas adds a miracle more:
quantum indeterminacy allows for it to be ideally enclosed, 
at the prescribed values of thermodynamic potentials,
in a sphere just small enough for the Bekenstein bound
to be attained.

The coefficient $C$ in the above relation $l^* = C \, \bar r$ 
is admittedly not very sharply defined;
this fits with the highly idealized situation we are considering
with photons in a ball with diameter just a little larger than
their own wavelength.
The message we get anyway, 
is that what at the end precludes
the Bekenstein bound from being violated
within the hydrodynamic regime
are bounds (\ref{From.12}), (\ref{Onthe.6}),
which is to say
quantum indeterminacy 
(\cite{IvanovVolovich, BoussoQuantum->Bekenstein} and \cite{PesC}).
We might still contemplate smaller balls with potentials
safely described by the hydrodynamic regime, in principle
violating the Bekenstein bound;
but in that case quantum indeterminacy would compel the gas in the balls 
to have values of the thermodynamic potentials different
from the prescribed ones, in the manner needed to protect the bound. 

Even if 
--the same as for relaxation times and viscosity/entropy--
the Bekenstein bound does not involve gravity,
(as can be envisaged by inspecting the constants in (\ref{BekensteinBound})),
by a historical accident we might say, its first derivation
was through the mentioned consideration of processes
involving black holes \cite{BekensteinSpecificEntropy}, 
and Schwarzschild black holes clearly 
attain the bound: 
$\frac{S}{E} = \frac{\pi R^2}{R/2} = 2 \pi  R$.
A full proof of the Bekenstein bound exists 
in quantum field theory \cite{Casini}.
It is based on a notion of entropy given by the difference
of actual entropy and the entropy of vacuum state;
this turns out to be finite (contrary to actual entropy
and vacuum entropy taken separately).
In this derivation the bound arises as the nonnegativity
of relative entropy between local density matrices describing
the vacuum state and the actual state reduced in the volume $V$. 
Similarly to the proof \cite{quantumBoussoProof} 
it applies when backreaction is weak,
thing that, when the circumstances are such that
the only gravity potentially relevant to the case in question 
is the self-gravity of
the body, which is the case of the Bekenstein bound,
means that strictly speaking
it applies to nearly Minkowski
spacetime \cite{Casini}.

It would be interesting to have a proof of the bound for nonnegligible
self-gravity (with the only constraint being that the elements 
which enter to define the bound
can be meaningfully defined).
A proof for these circumstances indeed exists \cite{PesE}, 
not in quantum field theory 
but working instead within the hydrodynamic approximation
and using the inequality (\ref{Onthe.6}) above.
A full proof in quantum field theory for strong self-gravity,
thus transcending also in this case the need of any microscopic
relation between energy and entropy densities,
would clearly be highly desirable.
In the meantime a quantum field theory proof of flat-spacetime relation 
(\ref{Onthe.6}) might also be welcome.
Indeed,
when joined with the proof \cite{PesE}, it would lead to a sort of 
QFT-based hybrid proof of the Bekenstein bound for generic
self-gravity.  

The same as for the viscosity/entropy bound,
also the Bekenstein bound 
might in principle be invalidated by a growing, ideally unboundedly, 
number of particle species as repeatedly noticed in literature.
Point is that we can increase the number of species in a system,
thus increasing its entropy, while keeping fixed its energy,
and we might expect this way to break the bound.
In the QFT proof \cite{Casini} it is shown 
that with that notion of entropy this can not happen.
The number $I$ of species affects indeed both the system
under consideration and the vacuum and the difference
of their entropies saturates for large $I$ allowing
the bound to be safe. 
 
Also in the approach we are following here
the species problem seems can be avoided.
This is basically the expression of nonnegativity of chemical potentials
when they include the mass energy of the particles.
If we go from a homogeneous system
to an inhomogeneous one with $I$ species
while keeping a same energy, pressure and temperature,
the Gibbs-Duhem relation goes from
$\rho = T s - p + \mu n$ 
to  
$\rho = T {\tilde s} - p + \sum_{i=1}^I \mu_i n_i$,
where $\tilde s$ is the entropy of the inhomogeneous system
and $\mu_i$, $n_i$ are the chemical potential and number density
of the species $i$ in it.
Then
${\tilde s } = s + \frac{1}{T} (\mu n - \sum_{i=1}^I \mu_i n_i)$,
and from $\mu_i \ge 0$ we get 
${\tilde s}_{max} = s + \frac{\mu n}{T}$
as the maximum entropy among all the inhomogeneous systems
at the assigned thermodynamic conditions.
This gives a corresponding characteristic length
${\tilde l}^*_{max} = \frac{1}{\pi} \, \frac{\tilde s_{max}}{\rho + p}
= \frac{1}{\pi T}$.
Now, the de Broglie wavelength $\lambda$ of a particle
of mass $m$ is (with all units)
$\lambda = \frac{h}{P} = \frac{c}{v} \lambda_{\gamma} > \lambda_{\gamma}$
where $\lambda_\gamma$ is the wavelength of a photon at same energy
(corresponding to the assigned temperature)
and $P$ and $v$ are particle's momentum and velocity.
This means that for no one of the species we have
$\lambda_i < \frac{1}{\pi T}$ where $\lambda_i$ is de Broglie
wavelength of species $i$ at temperature $T$.
Thus, since by definition $l > \lambda_i, \forall i$,
or at least $l > \langle \lambda_i \rangle$ where the latter
is some average over the $\lambda_i$'s,
we get $l > {\tilde l}^*_{max}$
and there is no violation of bound (\ref{Onthe.6})
relative to inhomogeneous systems (in (\ref{Onthe.6}), $\mu n$ 
is replaced by $\sum_{i=1}^{I} \mu_i n_i$)
and thus of the Bekenstein bound 
(see also \cite{PesE}).

\subsection{Why can the GCE bound be actually attained?}

This concludes the description of items seemingly distant at first,
yet strongly connected, with conditions (\ref{From.12}, \ref{Onthe.6}).
Let us come back to these conditions
in their relation with the generalized covariant entropy bound.
What we have seen so far is that these inequalities
are safe 
due to quantum indeterminacy 
and can be actually attained
by the most entropic systems (in specific geometric arrangements);
and that when these conditions are then joined to
Einstein's field equations,
we get the generalized covariant entropy bound 
(\ref{genBousso}) is preserved.

Now, two further intriguing aspects
appear to be worth noticing.
One
(the other is left to the next subsection)
is indeed  
the result of that paper \cite{PesG} which 
first aroused Paddy's curiosity.
It is the fact that in so doing, namely preserving
bound (\ref{genBousso}), 
Einstein's equations turn out to be
right what allows bound (\ref{genBousso})
to be exactly
attained by the most entropic systems 
in the most convenient setups
(namely, a layer of a thermal photon gas at temperature $T$ 
of thickness $l = \frac{1}{\pi T}$ 
with lightsheet given by orthogonal null geodesics emanating
from one of its faces 
crossing it with initially vanishing expansion,
i.e. a `plane' layer).
Point is that this is surprising because
we appear to have in principle no reason for such a perfect match.
Given Einstein's field equations,
the validity of the generalized bound (\ref{genBousso})
requires, in the hydrodynamic regime, 
a constraint in the local thermodynamic potentials,
which takes the form of inequality (\ref{Onthe.6});
there is no surprise in this.
What is surprising is that inequality (\ref{Onthe.6})
in specific circumstances can be attained,
while generically one might had expected it to be satisfied by far
in any circumstances.
Indeed, 
once the generalized bound (\ref{genBousso}) is accepted as true
(the factor 1/4 coming e.g. from the spherical entropy bound,
with no hint that the gap between black hole entropy and matter entropy
within an assigned spherical area can be made small, rather just the opposite), 
why has gravity to act in just that way 
that gives the generalized bound exactly attained 
in specific cases?
The effects of gravity might 
give the bound satisfied 
by orders of magnitude to spare
also for the most entropic systems and in the most
favorable circumstances,
why is that the bound fits in such a perfect manner instead?

There seems to be sort of coincidence or fine-tuning here
which might deserve further scrutiny.
It might suggest that gravity is driven by
matter dofs, or more precisely by all possible
matter dofs at the assigned $\rho+p$ \cite{PesG},
meaning this that we have to include also 
those that, 
at the assigned thermodynamic conditions,
are concealed 
in the mass of the particles forming the system.
We might
take this as something hinting 
towards a statistical origin of gravitational dynamics 
in the same vein as 
the various derivations
of field equations 
by thermodynamic arguments
or, before that,
as something further hinting to gravity
as an emergent phenomenon arising in an appropriate statistical limit
from a fundamental (and unknown) microscopic theory of gravity and matter,
and as such prone to a thermodynamic description 
as in \cite{PadI, PadJ, PadK} up to the point to allow 
the thermodynamic language to replace the geometric one
\cite{Pad08}.
This adds some elements to the big question
about the role of field equations in the process
of unveiling a quantum theory for gravity:
Are they a specific indication 
hinting to the precise underlying microscopic theory?
Or would they come about the same instead, as a pure statistical effect,
whichever this underlying theory is (somehow like   
the equation of state of a gas 
which happens can be calculated the same 
even irrespective of the fundamental description is taken to be classical
or quantum mechanical)?

\subsection{Unavoidable nonlocality}

The second aspect we would like to emphasize
is that the bound (\ref{genBousso}) adds a new
element, not contemplated 
by inequalities (\ref{From.12}, \ref{Onthe.6}),
and which can not be accounted for, 
as (\ref{From.12}, \ref{Onthe.6}) instead,
by quantum uncertainty,
this fact suggesting that quantum indeterminacy is 
definitely not the end of the story. 

To see what is this about, let us take
our plane layer of thermal photon gas
with thickness $l = \lambda = \frac{1}{\pi T}$
between faces $B$ and $B'$.
We know that the 
lightsheet
crossing the layer
with null geodesics emanating orthogonally 
from $B$ (thus with vanishing expansion)
with area $A_B$
just attains the bound
with some, in principle only very slightly smaller, exit area $A_{B'}$
(it is the geometry Fig. \ref{fig}, with matter forming a
very thin layer ($l$ extremely small) and the horizon playing the role 
of the lightsheet).
Let now slightly increase $T$, and then decrease $\lambda$ to a new value, 
and take as before $l = \lambda$, i.e. take a layer with thickness
equal to the new $\lambda$, then comprised between $B$ and a new
$B'$ nearer.
For this new layer, the generalized bound (\ref{genBousso})
will be again exactly attained, but this time with an $A_{B'}$ smaller
than before
(this means that in a smaller thickness
we get a smaller exit area;
we can readily understand this
if we consider that entropy density $s \sim T^3$
while the volume of the layer scales as $\lambda \sim T^{-1}$).
We see, 
the dependencies on temperature are such that
if we proceed this way ideally increasing $T$ at our wish,
we eventually
reach a temperature $\bar T$
at which the bound is still saturated by the associated layer,
but the exit area has become $A_{B'} = 0$.
This corresponds to a critical situation in which
the entropy of the gas `eats' all the entropy 
made available at start by the area $A_B$;
any further increase in $T$, and decrease in $\lambda$, 
is no longer allowed
as it would require the gas to have more entropy
than what available.  
This sets an upper limit 
$
S \le \frac{A_B}{4}  
$
to the entropy $S$
of the layer of a thermal photon gas 
we can build on the plane surface $B$ 
with thickness equal to 1 photon wavelength
\cite{PesH},
or equivalently the bound

\begin{eqnarray}\label{15.2}
s \, \le \, \frac{1}{4 \bar\lambda} \,
\Big[ = \frac{1}{4 \bar\lambda} \, \frac{k c^3}{\hbar G}\Big]
\end{eqnarray}
to entropy density,
where $\bar\lambda$
is (photon) wavelength corresponding to the critical temperature $\bar T$.

We can actually compute $\bar\lambda$.
Using the formula for entropy density $s$ of a thermal photon gas,
$
s = \frac{4 \pi^2}{45}  \, T^3,
$
from (\ref{15.2}) and $\bar\lambda = 1/(\pi {\bar T})$ we get
$
\frac{4}{45 \pi} (1/{\bar\lambda}^3) = \frac{1}{4} (1/{\bar\lambda})
$
which gives
$
{\bar\lambda} = \frac{4}{\sqrt{45 \pi}} \approx 0.3 \,
\Big[= 0.3 \sqrt{\frac{\hbar G}{c^3}} \Big]
= 0.3 \, l_p,
$
with $l_p = \sqrt{\frac{\hbar G}{c^3}}$ the Planck length,
i.e. $\bar\lambda = {\cal O}(l_p)$ as we would have expected.
Equation (\ref{15.2}) is then

\begin{eqnarray}\label{15.6}
s \, \le \, \frac{\sqrt{45 \pi}}{16} \, \frac{k}{l_p^3}
\, \approx \, 0.74 \, \frac{k}{l_p^3}. 
\end{eqnarray}
If the same microstates are counted in binary log
(to give information in bits), we get
$
\frac{s_{(2)}}{k} \le
\frac{\sqrt{45 \pi}}{16} (\log_2 {\rm e}) 
{\rm bits}/{l_p^3}$ 
$
\approx
1.07 \, {\rm bits}/l_p^3, 
$
where $s_{(2)}$ is entropy in base 2.

We see what is the new element introduced with the generalized bound
(\ref{genBousso}) not contained in the
bounds (\ref{From.12}, \ref{Onthe.6}):
it is the existence of a limit length 
$\bar\lambda = {\cal O}(l_p)$ which we cannot go below
when accounting of the dofs of any physical system. 
This is clearly connected with the fact that bound (\ref{genBousso})
relies on and manifests gravity, as can be envisaged from (\ref{15.2}) or 
from the expression of the bound (\ref{genBousso})
itself with all units set in place,
$S(L) \le 
\frac{A_B - A_{B'}}{4} \frac{k c^3}{\hbar G}$, 
which explicitly exhibits $G$.

In the absence of a finite $\bar\lambda$,
there would be no limit on how small the length scale could be 
when considering inequalities 
(\ref{From.12}, \ref{Onthe.6}).
$l^*$ for a thermal photon gas could become vanishingly small,
and yet inequality (\ref{From.12}) would always hold true
thanks to quantum indeterminacy.
Since in the hydrodynamic regime
the Bekenstein bound follows from inequalities 
(\ref{From.12}, \ref{Onthe.6}) with no extra input, 
this means that there is
no intrinsic need from it for a limit length scale,
at striking variance with what happens
with the generalized bound (\ref{genBousso}).

This 
has consequences also for the hydrodynamic approximation.
As far as say inequality (\ref{From.12}) for a thermal photon gas holds
true, and can be actually attained
without any limit on how small the thickness $l$ can be,
there is virtually no lower limit to the length scale 
of applicability of the
hydrodynamic approximation and 
thus to the local microscopic correspondence
between energy and entropy.
With a finite limit length $\bar\lambda$ however,
we can no longer resort to (\ref{From.12}) 
if we hypotetically go below a thickness $\bar\lambda$,
and the hydrodynamic approximation as well as 
the local microscopic correspondence between energy and entropy 
must at this scale necessarily break down.
We can thus see that at length scale $\bar\lambda$
nonlocality unavoidably enters the scene,
even if we carefully avoided at start 
any circumstances
requiring a nonlocal description.

Since as mentioned
the limit length $\bar\lambda$ appears
as soon as gravity is involved,
from what we just said we are lead to conclude that nonlocality
irreducibly accompanies gravity.
With gravity, quantum indeterminacy alone is not enough,
nonlocal correlations are also unavoidably required.  
This brings to further appreciate how convenient are the mentioned proofs
\cite{quantumBoussoProof} of the generalized bound
and \cite{Wall1, Wall2} of the generalized second law,
or \cite{Casini} of the Bekenstein bound, 
in that their scope extends to general nonlocal circumstances,
explicitly refraining from
any assumption about a local microscopic relation
between energy and entropy.
What the present discussion might add perhaps,
is a way to see first hand
that with gravity you necessarily get some form of nonlocality,
no matter how hard you struggle to keep off it. 
This apparently selects nonlocality as an unavoidable ingredient 
in the structure of spacetime.

\section{Spacetime and (a bit of) statistical mechanics}

\subsection{(Small-scale) nonlocality built in: Minimum-length metric}

What we have seen in the above adds 
but a small piece to
the gigantic mosaic of results
which show the need for a minimal length
when generically
combining general relativity and quantum mechanics
(a review can be found in 
\cite{Garay, Hossenfelder} and references therein).
In what we described,
the specific stress
has been on that, based on entropy bounds, 
nonlocality appears as a characteristic feature
accompanying gravity, 
shaping in particular the small-scale
texture of spacetime through the existence 
of a minimal length.
Next logical step is to try to make sense of this form
of irreducible nonlocality of geometry.

A natural way to implement this is to require
that no observable can distinguish between two events
when their classically-expected separation is small enough;
this might correspond indeed 
to the intrinsic impossibility to account
dofs below a certain separation scale,
this coming, as discussed above, 
from the entropy bound (\ref{genBousso}).   
The distance itself between events, in particular, 
should be such that it can not tell any better than
the length scale of this unavoidable nonlocality
when the two events are brought so near to each other
to classically coincide.  

What we ought to describe is a spacetime
whose geometry generically coincides on large scales
with the classical semi-Riemannian geometry 
of general relativity,
but deviates from it in the small scale,
very strongly
as soon as two events $p$, $P$ come to
be very near to each other,
giving a finite limit geodesic distance $L$
(in principle ${\cal O}(l_p)$,
but more generally 
we might think also larger) 
between them 
when $p\to P$.
%
At an effective level, 
one might want this transition to be smooth.

The generic necessity of such a modified description
of spacetime has been recognized long since.
Yet the specific mathematical characterization of it
has turned out to be 
not an easy task at all, 
not least because any metric-like tensor which converts
a separation into a distance ought to be divergent at a point
if it has to provide a finite result in the coincidence limit;
one has also to find a way to implement the limit distance
as a Lorentz invariant notion.
But above all,
there is no guidance 
(in absence of an agreed-upon quantum theory of gravity)
on which physical aspect should drive
the crossing of the classical-to-quantum interface 
in the small separations,
and e.g. signature changes ought
also to be contemplated 
possibly bringing spacetime
at the smallest scales to be Euclidean
(see e.g. \cite{WhiteWeinfVisse}).

Only quite recently
a modification of spacetime geometry along the lines above
has found
a solution, as something of a miracle,
in \cite{KotE}, \cite{Pad01} and, after further specification, 
in \cite{JaffinoKot}.
This has been done in a framework which formally
assumes that there is no change of signature
when going to the smallest scales.
As we will realize below, the procedure works 
equally well both in the Lorentzian and the Euclidean cases;
therefore the results in the Euclidean case
might be anyhow a useful description were
the metric at the smallest scales Euclidean.  
What is missing is the actual handling of a possible change
of signature while going to coincidence; 
this might be worth exploring,
for example along what suggested and developed  
in \cite{KotL, KotL_bis, SinghKothawala}.

In the aforementioned solution,
the nonlocality intrinsic to the existence of a limit 
geodesic distance $L$ 
is assumed to be captured by going from tensors to bitensors.
Of these,
the Synge's world function $Sy(p, P)$ \cite{Synge} stands out,
since when given between any two events in a 
geodesically convex region of spacetime 
it completely characterizes the metric properties of that region.  
The prescription is then to modify $Sy(p, P)$,
or the squared interval $\sigma^2(p,P) = 2 \, Sy(p, P)$,
between any two assigned events
(assumed to be one in a normal neighborhood of the other)  
to a new squared interval $S_L$ which is assumed to depend
on $\sigma^2$ alone, $S_L = S_L(\sigma^2(p, P))$.
The general framework is that equigeodesic surfaces for the ordinary
metric do result equigeodesic surfaces of the same nature
(spacelike, timelike, null) in the new metric,
down to the smallest scales.

To ensure that spacetime with the 
new squared interval biscalar $S_L$
has the metric properties just mentioned,
one requires that 
$i)$ $S_L(\sigma^2(p, P))$ approaches $\sigma^2(p, P)$ for large separations,
and 
$ii)$
$S_L(\sigma^2(p, P)) \to L^2$ when $p\to P$ along a spacelike geodesic,
and similarly $S_L(\sigma^2(p, P)) \to -L^2$ if along
a timelike geodesic 
(we use mostly positive signature for the metric $g_{ab}$).
With these positions,
the null cone from $P$ is unavoidably a discontinuity
surface for $S_L$ (we get $L^2$ when approaching it from one side
and $-L^2$ from the other);
the request that null equigeodesic surfaces according to the ordinary
metric be null also for the qmetric fixes $S_L = 0$ on the null cone
at any $p$ distinct from $P$.
Notice that the rule that for any given pair of events 
$S_L$ depends only on $\sigma^2$ applies unaltered
also to the $\sigma^2 = 0$ case.

$\sigma^2$ is clearly geometrical, it is a squared distance;
then,
since for any assigned $(P, p)$, $S_L$ depends only on $\sigma^2$, 
the prescription just given is also completely
geometrical, independent of the charts we may use
to map the manifold.
This in particular implies that whenever 
time- or space-separated events $P$ and $p$ are near enough
that they can be both described by a single local Lorentz frame,
based e.g. at $P$, $S_L$ is the same whichever is
the actual Lorentz frame we choose, i.e. it is locally 
a Lorentz invariant notion.

Clearly Synge's world function is related to the metric tensor, 
the main tool describing the metric properties of a manifold.
The relation between the two is best captured
by the following formula 
(written here in terms of $\sigma^2(p, P)$ instead of $Sy(p, P)$) 
\cite{PPV}

\begin{eqnarray}\label{Synge-metric-relation}
g^{ab}(p) \,\, 
\big(\partial_a^{|p} \sigma^2\big) \big(\partial_b^{|p} \sigma^2\big) 
\, = \, 4 \sigma^2 \, = \, 
g^{ab}(P) \,\, 
\big(\partial_a^{|P} \sigma^2\big) \big(\partial_b^{|P} \sigma^2\big), 
\end{eqnarray}
with
the superscripts in the derivatives denoting the events at which 
the derivatives are taken. 
As said, the nonlocality associated to the new squared intervals $S_L$
is assumed to be captured by bitensors. 
In particular we have to think
of a metric-like bitensor $q_{ab} = q_{ab}(p, P)$ 
(which usually goes under the name of
minimum-length metric or quantum metric or
simply qmetric) 
as replacing the metric tensor at $p$,
$g_{ab}(p)$, in providing the squared interval to $p$ (which we think of
as a field point) from $P$ (taken as base point). 
Clearly, $q_{ab}$ must be related to $S_L$ 
by a formula like (\ref{Synge-metric-relation}), i.e. 

\begin{eqnarray}\label{Synge-qmetric-relation}
q^{ab}(p, P) \,\,
\big(\partial_a^{|p} S_L\big) \big(\partial_b^{|p} S_L\big)
\, = \, 4 S_L,
\end{eqnarray}
where partial derivatives are taken with respect to
coordinates, which are the same as above. 

Because
the equigeodesic surfaces from $P$ in the ordinary metric
act also as equigeodesic surfaces of the qmetric,
any vector orthogonal to an  
equigeodesic surface in the ordinary metric at an event $p$,
and then tangent to the geodesic from $P$ according to the same metric,
is also tangent to a geodesic from $P$ according to the qmetric.
This brings with it that the same set of events 
which act as image of a geodesic curve from $P$
according to the ordinary metric are also image of a
geodesic according
to the qmetric, i.e. when we go to the qmetric
we map images of geodesics to images of geodesics;
what changes is when a given parameterization is affine.   

Our aim is now to find the expression of the qmetric
in terms of quantities relative to the ordinary metric.
Denoting with $t^a$ the ordinary unit tangent to the geodesics at $p$
($t^a = dx^a/ds$ with $s$ ordinary (taken positive) 
geodesic distance 
and $x^a$ the coordinates), 
the requirements $(i)$ and $(ii)$ above
lead to guess for $q^{ab}$ a form like \cite{Pad01, JaffinoKot}

\begin{eqnarray}\label{guess_qab}
q^{ab} = \frac{1}{A} \, g^{ab} 
+ \epsilon \, \Big(\alpha - \frac{1}{A}\Big) \, t^a t^b,  
\end{eqnarray}
($\epsilon \equiv g_{ab} t^a t^b = \pm 1$),  
with $A = A\big(\sigma^2(p, P)\big)$ a biscalar 
unspecified at this stage 
apart from having
$A\to \infty$ in the coincidence limit $p\to P$ and
$A\to 1$ for large separations,
and $\alpha = \alpha\big(\sigma^2(p, P)\big)$ another biscalar
designed to be responsible for giving a finite limit distance
for $p\to P$ and subject to the constraint $\alpha\to 1$ for diverging
separations.
Indeed,
we see that 
in this way 
(for any non vanishing, possibly diverging $\alpha$ when $p\to P$)
$q^{ab} \simeq \epsilon \, \alpha \, t^a t^b$
in the $p\to P$ limit, 
and 
$q^{ab} \simeq g^{ab}$ when separations are large.
From $q^{ac} q_{cb} = \delta^a_b$,
the covariant components of qmetric have the form \cite{Pad01, JaffinoKot}

\begin{eqnarray}\label{qabK}
q_{ab} = 
A \, g_{ab} + \epsilon \, \Big(\frac{1}{\alpha} - A\Big) t_a t_b,
\end{eqnarray}
with $t_a = g_{ac} t^c$.

Since (\ref{guess_qab}) can be recast as
$
q^{ab} = \frac{1}{A} h^{ab} + \epsilon \, \alpha \ t^a t^b,
$
with $h^{ab} = g^{ab} - \epsilon \, t^a t^b$ the metric transverse to $t^a$,
we see that $\alpha$ alone is involved 
in (\ref{Synge-qmetric-relation}) \cite{Pad01},
for as mentioned already, since $S_L$ depends on $\sigma^2$ alone, 
$\partial_a S_L$ is directed as $\partial_a (\sigma^2)$ 
and has then
no transverse component.
From 

\begin{eqnarray}
q^{ab} \, (\partial_a S_L) (\partial_b S_L)
\, = \, \epsilon \, \alpha \, (t^a \partial_a S_L)^2
\, = \, 4 S_L
\end{eqnarray}
one can get \cite{Pad01}

\begin{eqnarray}\label{alphaK}
\alpha = \frac{S_L}{\sigma^2} \, \frac{1}{S'^2_L},
\end{eqnarray}
with
$S'_L \equiv dS_L/d(\sigma^2)$.
Clearly $\alpha\to 1$ for diverging separations,
and, as also required, does not vanish (it diverges actually) for any $S_L$ 
whose $S'_L$ does not diverge when $p\to P$.

Another input is needed to fix $A$.
Here basic results from quantum gravity
come to the rescue.
In particular, quantum gravity effects
are expected to
change the coincidence limit 
of the Green's function of a free relativistic particle 
from $\sim 1/\sigma^2$ 
to $1/(\sigma^2 + \ell^2)$
with $\ell = {\cal O}(l_p)$
\cite{DeWitt, Paddy1985} (for review, 
\cite{Garay, Hossenfelder}). 
This,
which motivated in the first place to look at squared distances
(instead of e.g. to consider the metric itself)
when trying to set up a description of a quantum spacetime \cite{KotE},  
suggests to request the following:
the qmetric Green's function $G_{(q)}$
-meant as 
$G_{(q)}(\sigma^2) = G(S_L(\sigma^2))$
where $G(\sigma^2)$ is
the ordinary Green's function $G$-
do satisfy the same equation in terms
of the qmetric d'Alembertian $\Box_{(q)}$ 
as the ordinary Green's function $G$ does
for the ordinary $\Box$ \cite{KotE, Pad01, JaffinoKot}.
Equivalently,
if we instead {\it define} $G_{(q)}$ 
as what satisfies the equation with $\Box_{(q)}$,
we require $G_{(q)}(\sigma^2) = G(S_L(\sigma^2)$.
Since we are assuming that the new squared distances $S_L$
depend only on $\sigma^2$,
this condition has to be imposed on spacetimes in which 
the Green's function $G$ also depends
only on $\sigma^2$, 
such as Minkowski spacetime
or in all generality 
the maximally symmetric spaces \cite{JaffinoKot};
clearly the qmetric we get with the $A$ selected in this way 
will then apply to generic spacetimes. 
Adding to $(i)$, $(ii)$,
we have thus the following requirement \cite{JaffinoKot}:
iii) ${\Box}_{(q)} G_{(q)} = 0$ when 
 $\Box G = 0$ ($p \ne P$)
in all maximally symmetric spaces
all along the geodesic which from $P$ goes through $p$, 
with the qmetric Green's function taken to be
$G_{(q)}(\sigma^2) = G(S_L(\sigma^2))$.
In case of a locally Euclidean metric,
this is mirrored in taking
the $D$-dimensional Laplacian as replacing 
the d'Alembertian.

The magic of it is that
adding this third requirement completely 
and consistently 
(one gets the correct limits for $p\to P$ and
for diverging separations)
specifies $A$.  
What one obtains \cite{JaffinoKot} is (in general $D$-dim spacetime)

\begin{eqnarray}\label{AK}
A = \frac{S_L}{\sigma^2}
\Big(\frac{\Delta}{\tilde\Delta}\Big)^{\frac{2}{D-1}},
\end{eqnarray}
where

\begin{eqnarray}\label{vanVleck}
\Delta(p, P) = 
-\frac{\det\big[-\nabla_a^{|p} \nabla_b^{|P} \, \frac{1}{2} \sigma^2(p, P)\big]}
{\sqrt{g(p) g(P)}}
\end{eqnarray}
is the van Vleck-determinant biscalar 
(\cite{vVl, Mor, DeWA, DeWB}; see \cite{Xen, VisA, PPV})
($g$ is the (ordinary) metric determinant),
and the other biscalar  $\tilde\Delta$ is
$\tilde\Delta(p, P) \equiv \Delta({\tilde p}, P)$ where
$\tilde p$ is at $\sigma^2({\tilde p}, P) = S_L$ 
on the geodesic which goes through $P$ and $p$
on the same side as $p$. 
We can see that $A \to 1$ for diverging separations 
and, since $\Delta(p, P) \to 1$ when $p\to P$,
$A$ diverges in the coincidence limit, as required. 

Once $S_L = S_L(\sigma^2(p, P))$ is given,
formulas (\ref{guess_qab}) or (\ref{qabK})
equipped with $\alpha$ and $A$ from
(\ref{alphaK}) and (\ref{AK}),
completely define the qmetric with base $P$ at any $p$ 
in any normal neighborhood of $P$, 
except at events null separated from $P$.

\subsection{Minimum-length metric and null separations}

Consideration of null separations 
appears not entirely straightforward.
It is clear moreover that 
the examination of this case
critically depends
on whether the Lorentz structure is maintained
at the smallest scales.
In what follows, 
as already mentioned in the general description above,
we assume this is the case.

$\sigma^2$, and $S_L$ with it, are
identically vanishing along the null geodesic $\gamma$ from $P$,
and equation (\ref{Synge-qmetric-relation}),
which has been key to obtaining a limit length,
looses any track of the point $p$ 
to which it is applied;
it is thus quite not immediately clear what might it imply 
on $\gamma$ the general mapping $\sigma^2 \to S_L$.
The fact is that $\sigma^2$ and $S_L$ 
are both of no use as concerns the ability
to pick out specific events along $\gamma$.
In the impossibility to select $p$ along $\gamma$,  
how can we possibly give $q_{ab}(p, P)$?

Let consider at $P$ a local observer with velocity $V^a$,
and (uniquely) fix an affine parameter $\lambda$ along $\gamma$,
with $\lambda(P, P) = 0$, by requiring $V^a l^b g_{ab} = -1$
with $l^a = (d/d\lambda)^a$ (tangent to $\gamma$).
This quantity is clearly perfectly fit
for individuating events on $\gamma$
(in a Lorentz invariant way). 
Then the question is:
does the general mapping $\sigma^2 \to S_L$ imply
a $\lambda_{(q)} = \lambda_{(q)}(\lambda(p, P))$
to which $\lambda$ is mapped
and which plays the role of affine parameterization 
according to the qmetric?
How would it look like? What its relation with $S_L$?

Of great help for this is the fact that 
the given $\lambda$
can be thought of as (nonnegative) distance $l$ along the geodesic 
according to the observer $V^a$
\cite{VisA}.
Indeed,
$
l(p, P) = \int_0^\lambda -V^a l^b g_{ab} \, d\lambda' = \lambda(p, P), 
$
with $V^a$ parallel transported along $\gamma$.
Since both the time and space quadratic distances $\sigma^2$ 
are sent to $S_L$,
$l$ gets replaced in the qmetric by $\sqrt{S_L(l^2)}$,
this actually showing that the mapping to new quadratic distances 
$S_L$ induces a mapping from $\lambda$ to a new
parameter $\lambda_{(q)}$, 
which we have for consistency
to require as affine in the new metric, 
with

\begin{eqnarray}\label{lambdaq}
\lambda_{(q)}^2/\lambda^2 = S_L(l^2)/l^2.
\end{eqnarray}

Notice that $\lambda(p, P)$ is
a biscalar strictly speaking defined in the submanifold $\Gamma$
of codimension 1
swept by all null geodesics emerging from $P$ with $V^a l^b g_{ab} = -1$.
It can however be
extended infinitesimally off $\Gamma$. 
Any smooth extension can be described
thinking to geodesics
originating not only exactly from event $P$ with coordinates 
$X^a = (0, \vec{0})$ in frame $V^a$
but simultaneously at time $X^0=0$ from a
small ball $B$ of radius $r$ centered at ${\vec X} = {\vec 0}$.
All of them with $V^a l^b g_{ab} = -1$ and  
with $l^a$ at start at any point $\vec{X} \in B$ 
with spatial component 
in general slightly different, in an arbitrary fashion,
from the direction of $\vec{X}$.
We can then think of $\lambda(p, P)$, equipped with any smooth extension,
as a biscalar defined in a neighborhood of $\Gamma$.

What we are doing is
to take as $\lambda$ 
the proper time/space separations
according to observer $V^a$.
The requirements above of the time/space separation  cases
then actually translate into:
$I)$ $\lambda_{(q)}(p, P) \simeq \lambda(p, P)$ for $p$ diverging from $P$
 along $\gamma$
(meaning, for $\lambda(p, P)$ diverging),
and
$II)$  $\lambda_{(q)}(p, P) \to L$ for $p\to P$ along $\gamma$. 

Similarly to the nonnull separation case,
from these two requirements alone we can try to guess
a form
for $q^{ab}(p, P)$ for null separated $p$, $P$.
As we did in that case,
this turns out to be most conveniently done
singling out in the expression of $g_{ab}$
the part transverse to the vector tangent to the geodesic,
in present case to $l^a$. 
Since $l^a$ is null we need here an auxiliary null vector
to characterize the transverse metric.
We take it as
$
n^a \equiv V^a - \frac{1}{2} l^a.
$ 
It has normalization
$n^a l^b g_{ab} = -1$.
In terms of it the part $h_{ab}$ of $g_{ab}$
transverse to $l^a$ reads 
$
h_{ab} = g_{ab} + l_a n_b + n_a l_b,
$ 
and we can guess \cite{PesL, PesN}

\begin{eqnarray}\label{q^abnull}
q^{ab} &=&
\frac{1}{A_{\Gamma}} g^{ab} + \Big(\frac{1}{A_{\Gamma}} 
-\alpha_{\Gamma}\Big) 
(l^a n^b + n^a l^b)
\nonumber \\
&=&
\frac{1}{A_{\Gamma}} h^{ab} -\alpha_{\Gamma} \, (l^a n^b + n^a l^b),
\end{eqnarray}
where, analogously to the nonnull case, 
$A_{\Gamma} = A_{\Gamma}(\lambda(p, P))$  
with $A_{\Gamma}\to \infty$ when $p\to P$ along $\gamma$
and $A_{\Gamma}\to 1$ when $\lambda(p, P)$ diverges,
and $\alpha_{\Gamma} = \alpha_{\Gamma}(\lambda(p, P))$ 
responsible for giving a finite limit to $\lambda_{(q)}$ for $p\to P$
on $\gamma$
and subject to the constraint $\alpha_{\Gamma}\to 1$ for 
$p$ diverging from $P$ along $\gamma$.
Again from $q^{ac} q_{cb} = \delta^a_b$,
the covariant components are

\begin{eqnarray}\label{q_abnull}
q_{ab} = A_\Gamma \, g_{ab} + 
\Big(A_\Gamma - \frac{1}{\alpha_\Gamma}\Big) (l_a n_b + n_a l_b).
\end{eqnarray} 

We need an equation which replaces 
(\ref{Synge-qmetric-relation})
(which keeps being true of course),
in embodying the fact 
that geodesic distances $\sqrt{\epsilon \sigma^2}$
are mapped to new geodesics distances $\sqrt{\epsilon S_L}$,
read here distances $\lambda = l$ along the null geodesic $\gamma$
according to $V^a$
are mapped to new distances $\lambda_{(q)} = \sqrt{S_L(l^2)}$. 
This amounts to require
that in the qmetric
(i.e. in the requirements $(I)$ and $(II)$ above)
$\lambda_{(q)}$ is affine.
This reads \cite{PesN}

\begin{eqnarray}\label{affineq}
l^b_{(q)} \nabla_b^{(q)} l_a^{(q)} = 0,
\end{eqnarray}
where
$\nabla_a^{(q)}$ is the covariant derivative
in the qmetric 
\Big($\nabla_b^{(q)} l_a^{(q)} = 
\nabla_b \, l_a^{(q)} - \frac{1}{2} q^{cd} (-\nabla_d q_{ba} 
+ 2 \nabla_{\left(b\right.} q_{\left.a\right)d}) \, l_c^{(q)}$\Big) 
\cite{KotG},
and $l^a_{(q)} = dx^a/d\lambda_{(q)} 
= l^a \frac{d\lambda}{d\lambda_{(q)}}$ 
is the tangent
to the geodesic corresponding to the parameter $\lambda_{(q)}$
(for which one can easily verify that $q^{ab} \, l_a^{(q)} l_b^{(q)} =0$
in compliance with equation (\ref{Synge-qmetric-relation});
$l_a^{(q)} = q_{ab} \, l^b_{(q)}$).

Doing the calculations
\cite{PesN} 
shows that equation (\ref{affineq}) becomes

\begin{eqnarray}
\frac{d\lambda}{d\lambda_{(q)}} \, l_a \,
\frac{d}{d\lambda} \Big(\frac{d\lambda}{d\lambda_{(q)}} 
\, \frac{1}{\alpha_\Gamma}\Big) -
\Big(\frac{d\lambda}{d\lambda_{(q)}}\Big)^2
\Big(\frac{1}{\alpha_\Gamma} - A_\Gamma\Big) l^b \nabla_c l_b 
\, = \, 0.
\end{eqnarray}
Here the second term of l.h.s. vanishes since
$\partial_c (l^b l_b) = 0$ whichever is the extension
of a $l^a$ null off $\Gamma$. 
Then, the vanishing of first term requires
$\alpha_\Gamma = C \, \frac{d\lambda}{d\lambda_{(q)}}$ with $C$ a constant. 
$C$ is fixed to $C = 1$ from the constraint $\alpha_{\Gamma}\to 1$ for 
$\lambda \to \infty$.
The final expression
is then \cite{PesN} 

\begin{eqnarray}\label{alphanull}
\alpha_\Gamma = \frac{1}{d\lambda_{(q)}/d\lambda}.  
\end{eqnarray}
This can be compared 
with the expression we have for $\alpha$
for spacelike/timelike geodesics,
equation (\ref{alphaK}).
To this aim
the latter
can be conveniently recast in terms of affine parameters
of the geodesics,
and it then reads, as we can easily verify,

\begin{eqnarray}\label{alphaKrecast}
\alpha = \frac{1}{\big(ds_{(q)}/ds\big)^2},
\end{eqnarray}
where
$s \equiv \sqrt{\epsilon \sigma^2}$ 
and
$s_{(q)} \equiv \sqrt{\epsilon S_L}$
are (nonnegative) geodesic 
distances
for spacelike/timelike geodesics 
according respectively
to the ordinary metric and the qmetric. 

We see that equation (\ref{affineq}),
which further specifies 
equation (\ref{Synge-qmetric-relation}) 
in the null case, 
leaves $A_\Gamma$ unsettled,
analogously to what happens in the nonnull case.  
Clearly we need an equivalent of $(iii)$ 
for the null
separation case.
Here the problem is that the Green function $G(p, P)$ of the 
d'Alembertian diverges when $p$ and $P$ go to be null separated,
then all along $\gamma$. How to deal with this
when trying to require something like $(iii)$?

A way
is to take $\Box$ and $G$ not exactly on $\gamma$
but at points $p'$ slightly off $\gamma$,
thus time or space separated from $P$,
and then consider the limit $p'\to p\in \gamma$. 
The fact is that at any such $p'$,
in the limit $p'\to p$
the d'Alembertian of any function $f = f(\sigma^2)$
lends itself to be written as

\begin{eqnarray}\label{Box off gamma}
\Box f = \Big(4 + 2 \lambda \nabla_a l^a\Big) \, \frac{df}{d\sigma^2},
\end{eqnarray}
where $\lambda$ and $\nabla_a l^a$ are taken at $p$,
and analogously for the qmetric

\begin{eqnarray}\label{Box off gamma qmetric}
(\Box f)_{(q)} = \Big(4 + 2 \lambda_{(q)} \, \nabla_a^{(q)} l^a_{(q)}\Big) 
\, \frac{df_{(q)}}{dS_L(\sigma^2)},
\end{eqnarray}
\cite{PesL, PesN},
with $\sigma^2 = \sigma^2(p', P)$ along the (nonnull) geodesic $\gamma'$
through $p'$ from $P$. 
These relations allow the conditions on $f$ 
expressed in terms of $\sigma^2$
to be recast near $\gamma$ in terms of $\lambda$.

This is exactly what we need when trying to translate
the condition $(iii)$ above
for the d'Alembertian for space/time separations
to the case of null separations.
We take $f = G$ and consider $G(p', P)$ at $p'\notin\gamma$ (thus $G$ finite),
$p' \in \gamma'$ (nonnull),
with $p'$ going to approach $p \in\gamma$,
and require $(iii)$ along $\gamma'$.
From equations (\ref{Box off gamma}) and (\ref{Box off gamma qmetric}),
this translates into  
$III)$ 
$
\Big(4 + 2 \lambda_{(q)} \nabla_a^{(q)} l^a_{(q)}\Big) 
\, \frac{dG_{(q)}}{dS_L(\sigma^2)} = 0 
$
all along $\gamma'$ ($p' \ne P$),
when
$
\Big(4 + 2 \lambda \nabla_a l^a\Big) 
\, \frac{dG}{d\sigma^2} = 0
$
in the same,
in all maximally symmetric spaces,
with the qmetric Green's function taken to be
$G_{(q)}(\sigma^2) = G(S_L(\sigma^2))$.

We see this is equivalent to require
that along $\gamma'$ ($p' \ne P$)

\begin{eqnarray}\label{nullq31.3}
4 + 2 \lambda_{(q)} \nabla_a^{(q)} l^a_{(q)} = 0
\end{eqnarray}
when

\begin{eqnarray}\label{nullq31.4}
4 + 2 \lambda \nabla_a l^a = 0.
\end{eqnarray}
Now, computations show that 
equation (\ref{nullq31.3})
when expressed in terms of quantities defined 
in the ordinary metric becomes \cite{PesN} 

\begin{eqnarray}\label{nullq31.5}
4 + 2 \, \lambda_{(q)} \frac{d\lambda}{d\lambda_{(q)}} \nabla_a l^a 
+ (D-2) \, \lambda_{(q)} \frac{d\lambda}{d\lambda_{(q)}} 
\frac{d}{d\lambda} \ln A_\Gamma = 0.
\end{eqnarray}
If we then go to consider equation (\ref{nullq31.4})
in particular near $\tilde p \in\gamma$ 
with $\lambda({\tilde p}, P) = \lambda_{(q)}$  
(i.e. at $\tilde p'\in\gamma'$ approaching $\tilde p$),
we have there

\begin{eqnarray}\label{nullq31.6}
4 + 2 \, \lambda_{(q)} \, (\nabla_a l^a)_{|\lambda = \lambda_{(q)}} = 0.
\end{eqnarray}   
Using this in (\ref{nullq31.5}) we get

\begin{eqnarray}\label{nullq31.7}
-2 \, \lambda_{(q)} \, (\nabla_a l^a)_{|\lambda = \lambda_{(q)}}
+ 2 \, \lambda_{(q)} \frac{d\lambda}{d\lambda_{(q)}} \nabla_a l^a 
+ (D-2) \, \lambda_{(q)} \frac{d\lambda}{d\lambda_{(q)}}
\frac{d}{d\lambda} \ln A_\Gamma = 0.
\end{eqnarray}

Using a relation
which exhibits $\Delta$ as the ratio between the actual density of geodesics
and the density were spacetime flat
(here in the form specific to the null-congruence case) \cite{VisA}

\begin{eqnarray}\label{nullq31.8}
\theta = \nabla_a l^a = \frac{D-2}{\lambda} - \frac{d}{d\lambda} \ln \Delta
\end{eqnarray}
with $\theta$ expansion,
we see that equation (\ref{nullq31.7}) is equivalent to

\begin{eqnarray}
\frac{d}{d\lambda} \ln 
\Bigg[\frac{\lambda^2}{\lambda_{(q)}^2} 
\bigg(\frac{\tilde \Delta}{\Delta}\bigg)^\frac{2}{D-2} A_\Gamma \Bigg] = 0,
\end{eqnarray}
with $\tilde \Delta \equiv \Delta(\tilde p, P)$.
This fixes $A_\Gamma$ apart from a multiplicative constant
which is determined requiring $A_\Gamma \to 1$ for $\lambda\to \infty$.
The result is

\begin{eqnarray}\label{Anull}
A_\Gamma = \frac{\lambda_{(q)}^2}{\lambda^2}
\bigg(\frac{\Delta}{\tilde\Delta}\bigg)^{\frac{2}{D-2}}.
\end{eqnarray}
We see that $A_\Gamma$ 
diverges for $\lambda\to 0$ 
($\Delta(p, P) \to 1$ and $\tilde \Delta$ bounded
for $p\to P$)
coherently with the form chosen
for the qmetric.
Equations (\ref{q^abnull}) or (\ref{q_abnull}) together with
$\alpha_\Gamma$ and $A_\Gamma$ as given by (\ref{alphanull}) 
and (\ref{Anull})  
give the qmetric along any null geodesic from $P$.
$A_\Gamma$ is quite similar to
$A$ of equation (\ref{AK}) 
for time and space separations,
which we can rewrite
in terms of (nonnegative) geodesic distances 
$s$, $s_{(q)}$ as

\begin{eqnarray}\label{AKrecast}
A = \frac{s_{(q)}^2}{s^2}
\bigg(\frac{\Delta}{\tilde\Delta}\bigg)^{\frac{2}{D-1}},
\end{eqnarray}   
with
$\tilde\Delta$ the van Vleck determinant at $\tilde p$ with
$s(\tilde p, P) = s_{(q)}$.

The construction works once we are given 
an event $P$,
a null geodesic $\gamma$ from it
and an observer $V^a$.
This is at variance with the case of space and time separations,
for which there is no need of explicitly considering a local frame.
This stems from the fact that, 
given any two space- or time-separated events,
the squared interval along the geodesic connecting them
is uniquely determined, while the affine interval between
two null-separated events is not 
(it is defined up to a multiplicative constant). 
This complication seems then unavoidable, 
we have to live with it.

In ordinary spacetime once we are given an event $p$
it is uniquely associated to it  
the metric $g_{ab}$ at $p$.
By contrast,
in a spacetime with minimum length
the geometric object playing the same role
of characterizing the metric properties,
the qmetric $q_{ab}$,
is not defined when giving $p$ alone.
We have different nonequivalent geometric objects $q_{ab}$ at $p$ depending
on the choice of the base event $P$, and we need to specify
also the latter to fix the ambiguity, 
this fact embodying nonlocality. 
For null separations between $P$ and $p$
this geometric object
is assigned
only if we further
specify also an observer $V^a$,
i.e. we can assign to an event $p$
the qmetric only specifying in addition to the null-separated $P$
also an observer. 
Strictly speaking we should then write $q_{ab} = q_{ab}(p, P, V^a)$.
For example,
if we consider an assigned event $p\in\gamma$
we have $q^{ab} \approx g^{ab}$ according to an observer $V^a$ at $P$
which has $p$ at large space and time distance from $P$,
and  $q^{ab} \approx -\alpha_\Gamma (l^a n^b + n^a l^b)$
according to another observer $\tilde V^a$ 
which has $p$ at small space and time
distance from $P$.
Note that we do not have an (impossible) dependence of a geometric entity 
on the frame,
but we do have a dependence of the geometric entity we assign
on the frame.    

In the local frame of $V^a$ at $P$
if the affine parameter $\lambda$ is taken
to be space and time from $P$,
we see a certain structure in the qmetric at any  given $p\in\gamma$
as given by the just derived formulas.
The convenient parameterization $\lambda$ 
should be seen as instrumental
in extracting the structure of the qmetric;
and consistency demands that, once we have it,
this structure must be thought
as attached not to $\lambda$
but to the events $p$ to which $\lambda$ points:
when $p\in \gamma$ is at a small space and time from $P$ according
to $V^a$, we see the effects of a limit length,
regardless of any parameterization we may choose on $\gamma$.

If we change the local frame, 
in the new frame $\tilde V^a$ at $P$
we see that same structure,
but now in terms of space and times according to $\tilde V^a$,
and again irrespective of any parameterization on $\gamma$,
but depending only on the events $p$.
What happens then 
is that whichever is the local frame we are in
we observe a same structure
associated to the existence of a limit length.
In this sense we have Lorentz invariance of the
construction: the measured local structure of the qmetric
around an event $P$ (events null-separated from $P$ included)  
is the same according to every local observer at $P$;
with this we mean that the geometric object we assign to
compute distances is the same,
exactly as it happens in ordinary spacetime
in which this same object according to any observer
is invariably the metric $g_{ab}$.

Since the use of $q_{ab}(p, P)$ 
is in computing distances from $P$,
and we know from the beginning that they vanish on $\gamma$
(keeping Lorentz intact down to the smallest scales),
it might seem that all this effort about $q_{ab}$ on $\gamma$ null
is no big deal after all. 
We have to consider however that the specific form
of $q_{ab}$ on $\gamma$ null is relevant for a number of topics,
like e.g. to characterize the geometry of the $(D-2)$ spatial surface
to which $\gamma$ is orthogonal,   
or, if $\gamma$ is a member of a congruence of geodesics, 
to characterize the geometric properties of the congruence.

\subsection{`Points' with finite area}

With the given specifications for the null case,
what we have obtained for null separations,
when joined to the nonnull-separation results, 
allows to characterize the qmetric
along any congruence of geodesics emerging from an event $P$
in any local frame at $P$,
being the congruence timelike, spacelike or null.
This opens the way to compute possible
modifications to the Raychaudhuri
equation due to the existence of a minimal length.
But before that,
let us consider a very basic feature accompanying the qmetric
concerning the metric properties of the space transverse
to the direction along which the qmetric is taken.
It turns out that `areas' on equigeodesic surfaces
at $p$ do not vanish 
in the coincidence limit $p\to P$ 
and approach instead a finite value \cite{Pad06}. 

One might naively think this should be expected
in a spacetime endowed with a limit on distances.
But,
to appreciate that this might not be at all a trivial matter 
attention should be payed to that the areas we are talking
about are taken at $p$, not $P$.
The prescriptions we have given in $(i)$-$(iii)$ and $(I)$-$(III)$,
and which define the qmetric,
refer to affine distances from $P$.
One has no reason then e.g. to expect that the distance between two events
$p$ and $p'$ both on the same geodesic from $P$ do not vanish as evaluated
from $P$ (i.e. with the qmetric based at $P$)
when both approach $P$;    
one would expect this, and
as a matter of fact the qmetric volumes at $p$ 
do vanish in the limit $p\to P$ \cite{Pad06}
(in particular how quickly the volume approaches 0
can be used to infer
the dimensionality of qmetric;
taking a ball of radius $s$ 
in $D$-dimensional qmetric Euclidean space, 
its volume $V_D$ 
approaches 0 as $V_D \sim s^2$ when $s\to 0$ 
showing that the Euclidean space
(and, one might wonder, also the physical space) 
is effectively 2-dim
at the smallest scales \cite{PCK},
a result confirming several others 
concerning dimensional reduction
in quantum gravity
starting with \cite{AJL1, AJL2};
\cite{Carlip1, Carlip2} for review).    
It is then quite surprising and tricky 
that in the coincidence limit
finite areas do appear; 
besides, they show up orthogonally to the geodesic.

To see how this comes about we might follow the original
derivation \cite{Pad06} which was given in the Euclidean case
and then continued back to Lorentz to give $D-1$ limit areas
orthogonal to spacelike and timelike geodesics. 
But let us arrive here to the same results
in a slightly different way.
We choose to stay in Lorentz and to use,
beside equation (\ref{qabK}) for the qmetric for spacelike/timelike geodesics,
equation (\ref{q_abnull}) for null geodesics.
This brings 
to obtain with a similar procedure
the coincidence limits for orthogonal areas 
both in the spacelike/timelike and in the null case. 

The area of a small portion at $p$ of the equigeodesic
hypersurface $\Sigma_p$ of timelike or spacelike geodesics
from $P$ 
is according to the qmetric
$d^{D-1} a_{(q)} =
\sqrt{- \epsilon h_{(q)}} \, d^{D-1} a,
$
with $d^{D-1} a$ the area in the ordinary metric,
and $h_{(q)}$ the determinant of the transverse qmetric
$h_{ab}^{(q)} = q_{ab} - \epsilon \, t_a^{(q)} t_b^{(q)}$
\Big($t^a_{(q)} = (d/ds_{(q)})^a$, and $t_a^{(q)} = q_{ab} \, t^b_{(q)}$\Big). 
From (\ref{qabK}) one can verify that this gives
$
h_{ab}^{(q)} = A \, h_{ab}, 
$
and then

\begin{eqnarray}\label{orth_area_s/t}
d^{D-1} a_{(q)} = A^{\frac{D-1}{2}} d^{D-1} a
\end{eqnarray}
\cite{KotG}.

For light rays $\gamma$,  we have 
$
d^{D-1} a_{(q)} =
\sqrt{- \epsilon h_{(q)}} \, d\lambda_{(q)} \, d^{D-2} a
= d^{D-2} a_{\gamma (q)} \, d\lambda_{(q)}.
$
Here
$d^{D-2} a_{\gamma (q)}$ is the qmetric volume element 
of the ($D-2$)-space transverse to $\gamma$,
and $d^{D-2} a_\gamma$ the area of the same
according to the ordinary metric.
The transverse metric is
$ 
h_{ab}^{(q)} = q_{ab} + l_a^{(q)} n_b^{(q)} +  n_a^{(q)} l_b^{(q)}
$
\Big($n_a^{(q)} = \frac{d\lambda_{(q)}}{d\lambda} \, n_a$,
from requiring $q^{ab} \, l_a^{(q)} n_a^{(q)} = -1$\Big).
From (\ref{q_abnull}) one finds \cite{PesN},
analogously to the nonnull case,
$
h_{ab}^{(q)} = A_\Gamma \, h_{ab}
$
and

\begin{eqnarray}\label{orth_area_null}
d^{D-2} a_{\gamma (q)} = A_\Gamma^{\frac{D-2}{2}} d^{D-2} a_\gamma.
\end{eqnarray}

If we now follow towards $P$
the geodesics selected by the assigned small area,
the ordinary area elements intercepted on the
equigeodesic surface decrease with decreasing $s$ or $\lambda$
going to vanish in the $p\to P$ limit, 
while in the qmetric, from the expressions (\ref{AKrecast})  
and (\ref{Anull}) for $A$ and $A_\Gamma$ respectively, 
we definitely have in the same limit

\begin{eqnarray}\label{finite area s/t}
\Big(d^{D-1} a_{(q)}\Big)_0 
&\equiv&
\lim_{p\to P} d^{D-1} a_{(q)}
\nonumber \\
&=&
L^{D-1} \, \frac{1}{\Delta_L} \, d^{D-1}\eta 
\end{eqnarray}
and

\begin{eqnarray}\label{finite area null}
\Big(d^{D-2} a_{\gamma (q)}\Big)_0 
&\equiv&
\lim_{p\to P} d^{D-2} a_{\gamma (q)}
\nonumber \\
&=& 
L^{D-2} \, \frac{1}{\Delta_L} \, d^{D-2}\theta 
\end{eqnarray}
the former applying to spacelike/timelike geodesics
and the latter to light rays (\cite{Pad06} and \cite{PesN}).
Here $\eta^i$, $i = 1, ..., D-1$, and
$\theta^A$, $A = 1, ..., D-2$ are 
coordinates labeling the geodesics
($d^{D-1}\eta$ and $d^{D-2}\theta$ are then constant 
while $p$ approaches $P$) in orthogonal directions,
with $s \, d\eta^i$ being geodesic distances,
and $\lambda \, d\theta^A$ being distances according to observer $V^a$.
$\Delta_L \equiv \Delta(\bar p, P)$ with $\bar p$ the event
at $s(\bar p, P) = L$ on timelike/spacelike geodesics and
at $\lambda(\bar p, P) = L$ on light rays.
$\Delta_L$ is bounded for fixed $L$
(except for pathological circumstances; generically, $\Delta_L \approx 1$
for $L$ small; we will come back to this later),
and this shows that the two limiting areas (\ref{finite area s/t}) and
(\ref{finite area null}) are not zero.

\subsection{Raychaudhuri's equation gets modified: No focal points}

Clearly the existence of a finite limit area
can be expected to impact on the behavior of congruences
of geodesics when dealing with singularities
or, before that,
even simply with focal points.
Indeed the consideration as we did above 
of a congruence of geodesics emerging from $P$,
selects the event $P$ as a focal point of the congruence.
With no extra effort the results can also be applied in reverse direction 
since
reversing the sign of the affine parameters $s$ or $\lambda$
allows to read the results relative to a congruence
emerging from $P$ as relative to a congruence converging to $P$.
Let us consider then congruences of timelike, spacelike 
or null geodesics
emerging from an event $P$
and ask ourselves how their geometry appears in the qmetric,
considering specifically their expansion 
$\theta = \nabla_a t^a$, 
or $\theta = \nabla_a l^a$ in the null case.
This has been studied in \cite{KotG, PesM, ChaD}.
For spacelike and timelike geodesics 
from the expression 
of the qmetric covariant derivative \cite{KotG}
one finds that

\begin{eqnarray}\label{thetaq s/t}
\theta_{(q)} &\equiv& 
\nabla_a^{(q)} t^a_{(q)}
\nonumber \\
&=& 
\sqrt{\alpha} \, 
\Big[\theta + (D-1) \, \frac{d}{ds} \ln\sqrt{A}\Big],
\end{eqnarray}
and for light rays

\begin{eqnarray}\label{thetaq null}
\theta_{(q)} &\equiv& 
\nabla_a^{(q)} l^a_{(q)}
\nonumber \\
&=& 
\alpha_\Gamma \, 
\Big[\theta + (D-2) \, \frac{d}{d\lambda} \ln\sqrt{A_\Gamma}\Big].
\end{eqnarray}
On using the expressions for 
$\alpha_\Gamma$,
$\alpha$, 
and $A_\Gamma$, 
$A$, 
as given in terms of affine parameters
by equations 
(\ref{alphanull}, \ref{alphaKrecast})
and (\ref{Anull}, \ref{AKrecast}), 
and exploiting relation (\ref{nullq31.8}) for null congruences,
as well as its homologous \cite{VisA}

\begin{eqnarray}\label{nullq35.3}
\theta = 
\frac{D-1}{s} - \frac{d}{ds} \ln \Delta
\end{eqnarray}
for congruences of timelike and spacelike geodesics,
the expansion in the qmetric turns out to be

\begin{eqnarray}\label{thetaq s/t recast}
\theta_{(q)} 
\, \, = \, \,
\frac{D-1}{s_{(q)}} - \frac{d}{d{s_{(q)}}} \ln \tilde\Delta
\, \, = \, \,
\theta_{|s = s_{(q)}} 
\end{eqnarray}
and

\begin{eqnarray}\label{thetaq null recast}
\theta_{(q)} 
\, \, = \, \,
\frac{D-2}{\lambda_{(q)}} - \frac{d}{d{\lambda_{(q)}}} \ln \tilde\Delta
\, \, = \, \,
\theta_{|\lambda = \lambda_{(q)}} 
\end{eqnarray}
respectively for spacelike/timelike and null cases \cite{ChaD}. 
Comparing these 
with equations (\ref{nullq31.8}) and (\ref{nullq35.3})
which connect the expansion and van Vleck determinant
in the ordinary metric,
we see that a most direct way to get the expansion $\theta_{(q)}$
in the qmetric is just to replace in that equations
the affine parameters
$s$ and $\lambda$ with $s_{(q)}$ and $\lambda_{(q)}$,
and the van Vleck determinant $\Delta$ with $\tilde\Delta$.
Taking equations (\ref{thetaq s/t recast}) and (\ref{thetaq null recast})
as defining the van Vleck determinant $\Delta_{(q)}$ in the qmetric
-analogously to equations (\ref{nullq31.8}) and (\ref{nullq35.3})
which by integration define $\Delta$ in terms of $\theta$ and 
$\lambda$,  $s$ in the ordinary metric-
we get $\Delta_{(q)} = \tilde\Delta$, namely

\begin{eqnarray}
  \Delta_{(q)}(p, P)
=
  - \frac{1}{\sqrt{g(\tilde p) \, g(P)}} \,
  \det \bigg[- \nabla_a^{|\tilde p} \, \nabla_b^{|P} \,
    \frac{1}{2} \sigma^2(\tilde p, P)\bigg],
\end{eqnarray} 
with $\tilde p$ at $\sigma^2(\tilde p, P) = S_L$ on
spacelike/timelike geodesics, and at $\lambda(\tilde p, P) = \lambda_{(q)}$
on null geodesics. 

We can take $\theta_{(q)}$ from equations 
(\ref{thetaq s/t recast}) and (\ref{thetaq null recast}) 
and consider
the limit $(\theta_{(q)})_0$ we get when $p\to P$.
We have

\begin{eqnarray}
  \big(\theta_{(q)}\big)_0
  &=& 
  \frac{D-\delta}{L} - \frac{d}{dL} \ln \Delta_L,
\end{eqnarray}
with $\delta = 1$ (2) for spacelike/timelike (null) geodesics.

To make further statements about
$(\theta_{(q)}\big)_0$ we need knowledge of $\Delta_L$.
Expressions for $\Delta(p, P)$ are known through expansions in powers
of affine intervals $s$, $\lambda$ \cite{Xen}.
One finds that \cite{VisA}, \cite{PPV}  

\begin{eqnarray}\label{expansion}
\Delta(p, P) = 1 + \frac{1}{6} \, \ell^2 R_{ab} \, v^a v^b + {\cal O}(\ell^3), 
\end{eqnarray}
with $\ell = s, \lambda$  and $v^a = t^a, l^a$.
The part ${\cal O}(\ell^3)$ contains, among others, 
terms with any power of the quantity 
$\ell^2 R_{ab} v^a v^b$ and derivatives of any order of it.
The possibility for it to be
really negligible depends critically
on the value of $\ell^2 R_{ab} \, v^a v^b$ and its derivatives. 
From the mentioned geometrical meaning of van Vleck determinant
as ratio between the actual density of geodesics to that
were spacetime flat \cite{VisA},
we have that the sum in (\ref{expansion}) is 
not diverging provided that $p$ 
is close enough to $P$
to be not possible to be its conjugate.
At $\ell = L$ equation (\ref{expansion}) gives

\begin{eqnarray}\label{expansionL}
\Delta_L = 1 + \frac{1}{6} \, L^2 R_{ab} \, v^a v^b + {\cal O}(L^3), 
\end{eqnarray}
from which,

\begin{eqnarray}\label{thetaq_0}
\big(\theta_{(q)}\big)_0 \, = \,
\theta_{|\ell =L}
\, = \,
\frac{D-\delta}{L} - \frac{1}{3} \, L \, R_{ab}\, v^a v^b + {\cal O}(L^2)
\end{eqnarray}
\cite{ChaD}.

At generic conditions
$L^2 R_{ab} \, v^a v^b \ll 1$.
This gives $\Delta_L \approx 1$ 
and $\big(\theta_{(q)}\big)_0 \approx \frac{D-\delta}{L}$ finite.
Generically then,
caustics can not be formed in 
a qmetric spacetime, whichever is the geometry of the congruence
\cite{ChaD}.
However, at increasing curvature,
meaning at increasing $R_{ab}\, v^a v^b$,
$\theta_{|\ell =L}$ decreases
and $\Delta_L$ has then to increase, 
and actually diverges
for $R_{ab}\, v^a v^b$ large enough to have the event $\bar p$
(which is defined by $\ell({\bar p, P}) = L$)  
conjugate of $P$, i.e. $\theta_{|\ell =L} = 0$.
At these so huge values of $R_{ab}\, v^a v^b$,
which we might think can be found near singularities
of the ordinary metric,
$\Delta_L$ diverges and $(d^{D-1}a_{(q)})_0$ and
 $(d^{D-2}a_{\gamma \, (q)})_0$ 
of equations (\ref{finite area s/t}) and (\ref{finite area null})
can become in principle 0.
In these circumstances the construction
of the qmetric would hardly be applicable
since we can no longer have the crucial prerequisite
of the points $p$ and $\bar p$ to be in a normal neighborhood of $P$
(and thus have a unique geodesic connecting $P$ and $p$).

There is something artificial however in such unboundedly high values
of $L^2 \, R_{ab}\, v^a v^b$,
in the sense that if we are really given a spacetime which 
endows a lower limit length $L$,
$R_{ab}\, v^a v^b$ can arguably never diverge.    
Indeed, let us consider circumstances 
in which in ordinary spacetime
unboundedly high values of $R_{ab}\, v^a v^b$
actually do develop.
As a prototypical example we may take the collapse towards $P$ of a thin shell
of matter with exact spherical symmetry on a otherwise flat
Minkowski background.
It is clear that, as soon as $p$ taken on the shell
becomes nearer and nearer to $P$,  
arbitrarily high
values of $R_{ab}\, v^a v^b$ can be reached;
indeed these are circumstances of a classically blatant 
singularity formation.   
Yet, all along the path to $P$, 
never $p$ becomes conjugate of $P$
nor such becomes $\tilde p$ at $\ell(\tilde p, P) = \sqrt{\epsilon S_L}$,
and the qmetric construction does apply all the way.
Now,
in the considered circumstances
the qmetric prescriptions
demand that for $p$ at any $\ell$
we consider $\tilde p$ on the same geodesic
at the less evolved spacetime
$\ell(\tilde p, P) = \sqrt{\epsilon S_L}$.
This means that at coincidence,
i.e. at the most extreme conditions, 
we have to look
at the slightly antecedent situation corresponding to $\ell(\bar p, P) = L$,
at which everything is regular and the stress energy tensor, 
and thus the Ricci tensor, is finite.
We see that in all this, beside a bounded Ricci tensor, we have avoidance of 
classically-blatant singularity formation \cite{ChaD};
further investigation might be worth doing, exploring 
singularity formation avoidance of the qmetric
in more general settings.

Of help for this should be  
the qmetric Raychaudhuri equation

\begin{eqnarray}\label{Raychaudhuri basic}
\frac{d\theta_{(q)}}{d\ell_{(q)}} =
- \frac{1}{D-\delta} \, \theta_{(q)}^2 
- \sigma_{ab}^{(q)} \sigma_{(q)}^{ab} - R_{ab}^{(q)} \, v^a_{(q)} v^b_{(q)},
\end{eqnarray}
$l_{(q)} = s_{(q)}, \lambda_{(q)}$ 
($\sigma_{ab}$ is shear; no twist, from hypersurface orthogonality
of the congruence of geodesics emerging from $P$)
\cite{ChaD}.
In view of equations 
(\ref{thetaq s/t recast}-\ref{thetaq null recast}) it takes the form

\begin{eqnarray}\label{R.2}
-\frac{D-\delta}{\ell_{(q)}^2} - \frac{d^2}{d\ell_{(q)}^2} \ln\tilde\Delta 
&=&
- \frac{1}{D-\delta} \, 
\Big(\frac{D-\delta}{\ell_{(q)}} - \frac{d}{d\ell_{(q)}} \ln\tilde\Delta\Big)^2 
- \sigma_{ab}^{(q)} \sigma_{(q)}^{ab} - R_{ab}^{(q)} \, v^a_{(q)} v^b_{(q)},
\nonumber \\
\frac{d^2}{d\ell_{(q)}^2} \ln\tilde\Delta 
+ 
\frac{2}{\ell_{(q)}} \, \frac{d}{d\ell_{(q)}} \ln \tilde\Delta
- \frac{1}{D-\delta} \,  \Big(\frac{d}{d\ell_{(q)}} \ln\tilde\Delta\Big)^2
&=&
\sigma_{ab}^{(q)} \sigma_{(q)}^{ab} + R_{ab}^{(q)} \, v^a_{(q)} v^b_{(q)},
\end{eqnarray}
from which the coincidence limit of the r.h.s. can be extracted \cite{ChaD}.

Again concerning singularity formation, 
an interesting line of research
put forward recently
is their consideration in field-space 
rather than solely in spacetime \cite{CKK1, CKK2}.
It has been shown indeed (explicitly
in a cosmological setting)
that certain spacetime singularities
are not such in field-space and give place
to observables which are well-defined \cite{Kamenshchik et al, CKK3}.
One aspect worth investigating at this regard
might be if and how the fields are affected by the existence
of a limit length on the geometric side.

\subsection{Einstein's equations as a statistical-mechanical result}

The existence of a finite area at coincidence \cite{Pad06},
is a key prediction of the qmetric 
(equations (\ref{finite area s/t}) and (\ref{finite area null})).
It describes spacetime at the smallest scales
as possessing a structure, and tells something about the latter;
in particular it is suggestive of
elementary areas as what are made up
events.  
This structure hints to 
dofs associated with it,
which clearly we ought to think of as proper of spacetime itself.
This can tentatively offer a way to capture, 
in a very basic manner,
some features of the microscopics of the gravitational field,
this providing a basis for some statistical-mechanical description 
of the latter,  
even though in absence of a fully convincing quantum theory of gravity
and, moreover, irrespective of the specific form it might or will take. 
The meaning would be to get some specific results from the mere existence
of a minimal length (a generic prediction, the latter, shared by many quantum
approaches to gravity); these results are generic, they do not select any
theory in particular, but act as constraints on viable theories.
In addition they might offer some hints 
on how to conveniently shape or think of a quantum theory of gravity.  

Generically, when considering the states we can assign to two
`adjacent' (in some sense better defined below) 
independent events $P$ and $P'$, 
we expect their total number is the product of the states 
constituting $P$ and $P'$ separately.
Looking at equations (\ref{finite area s/t}), (\ref{finite area null}),
we see that the micro-based entities that are associated to events
$P$ and $P'$ sum up instead, since the total limit area corresponding
to the two is clearly the sum of the two areas.
This consideration suggests to regard the quantities
to the r.h.s.'s of (\ref{finite area s/t}) and (\ref{finite area null})
as expressing numbers of dofs
(kind of elementary dofs or `atoms' of spacetime \cite{Pad06, Pad10, Pad12}), 
not numbers of states.
Inspecting the form of the r.h.s.'s 
we are led moreover to interpret 
the quantity $1/\Delta_L$ as (proportional to) the number $N_G$ 
of gravitational dofs 
building up the event $P$ in the direction $v^a$,
to mean the number of dofs in area $L^{D-1}$ (area $L^{D-2}$ for $v^a = l^a$) 
transverse to $v^a$. 
We can write

\begin{eqnarray}\label{nullq43.1}
N_G \, = \, N_G(x, v)
\, = \, \big(N_G\big)_0 \, \frac{1}{\Delta_L}
\, = \, C  \, \frac{1}{\Delta_L},
\end{eqnarray}
\cite{Pad06},
where $x$, $v$  are short for coordinates $x^a$
and tangent vector $v^a$ at $P$,
and $(N_G)_0$ is the same as $N_G$ but for Minkowski.
$(N_G)_0$ can not depend on $x$, $v$ due to the symmetries of Minkowski,
and
to simplify notation in the last equality
we denote it as $(N_G)_0 \equiv C$ with $C$ being a positive
number, independently defined (and thus different) in each set 
of geodesics from $P$, i.e. if timelike, spacelike or null.
Using the series expansion (\ref{expansionL}) for $\Delta_L$ 
this can be expressed as 

\begin{eqnarray}\label{expansion density G}
N_G \, = \, C \, 
\Big(1 - \frac{1}{6} \, L^2 R_{ab} \, v^a v^b\Big) + {\cal O}(L^3).
\end{eqnarray}
From the mentioned geometrical meaning
of van Vleck determinant as 
ratio of densities of geodesics emanating from an event, 
we can geometrically view $N_G$ as $C$ ($(N_G)_0$) times
the ratio between the density of geodesics emerging from $P$
were spacetime flat and the density in actual spacetime \cite{Pad06}.

Having this,
we can consider beside gravitational dofs also 
matter dofs $N_m$ in $L^{D-1}$ (or $L^{D-2}$ for $v^a = l^a$) 
associated to matter entropy
at the coincidence limit.
With both,
we could then proceed to
describe the equilibrium configuration they eventually form
as what we get from extremizing total entropy 
or total number 
of microscopic configurations \cite{Pad06}
according to a principle of maximum entropy \cite{Jaynes01, Jaynes02}.
This has been investigated in \cite{Pad06, Pad10, Pad12, Pad16, PesK}
(for review, \cite{Pad20}),
in particular concerning the possibility to arrive this way
to field equations.
It turns out that this is indeed possible, and 
in the limit $L^2 R_{ab} \, v^a v^b \ll 1$ 
one gets
Einstein's field equations with cosmological constant,
the latter arising as an integration constant.

In $D=4$ the extremization of entropy leads to the
condition

\begin{eqnarray}\label{nullq41.3}
-\frac{1}{6} \, C \, L^2 \, R_{ab} v^a v^b + L^4 \, T_{ab} v^a v^b \, = \,
\lambda(x) \, g_{ab} v^a v^b,
\end{eqnarray}
where the second term in the l.h.s. comes from matter dofs
($N_m = L^4 \, T_{ab} v^a v^b \, 
[ = L^4 \, \frac{1}{\hbar c} \, T_{ab} v^a v^b]$ ; 
this, in case e.g. of $v^a$ timelike,
from $N_m = \delta Q/T$
with energy $\delta Q$ in $L^3$ given by  
$\delta Q = L^3 \, T_{ab} v^a v^b$ as probed along the geodesic,
at temperature $T \approx 1/L$, this coming from $\delta Q$ 
having to be in the volume $L^3$ \cite{PesK})  
and the extremization is taken indifferently in the sets
of spacelike or timelike or null tangent vectors $v^a$
at $P$. $\lambda$ is arbitrary, function of $x$ only not of $v$.
In the case of light rays, $v^a = l^a$,
this gives 

\begin{eqnarray}\label{nullq41.4}
-\frac{1}{6} \, C \, L^2 \, R_{ab} l^a l^b + L^4 \, T_{ab} l^a l^b \, = \, 0,
\end{eqnarray}
which implies

\begin{eqnarray}\label{nullq41.5}
-\frac{1}{6} \, C \, L^2 \, R_{ab} + L^4 \, T_{ab}  
\, = \, f(x) \, g_{ab},
\end{eqnarray}
with $f$ function of $x$.
Here, 
$\nabla_b {G^b}_a = 0$ joined with $\nabla_b {T^b}_a = 0$    
implies 
$
-\frac{1}{12} C L^2 \partial_a R = \partial_a f,
$
which gives
$
f =  -\frac{1}{12} C L^2 R + {\rm const}.
$
Writing 
$
f + \frac{1}{12} C L^2 R = {\rm const} \equiv \frac{1}{6} C L^2 \Lambda
$
with $\Lambda$ independent from $x$,
one easily verifies that equation (\ref{nullq41.5})  
becomes
$
\frac{1}{6} C L^2 
\Big(R_{ab} - \frac{1}{2} R g_{ab} + \Lambda g_{ab}\Big) =
L^4 T_{ab},
$
that is

\begin{eqnarray}\label{nullq41.6}
G_{ab} + \Lambda \, g_{ab} \, = \, 
\frac{6}{C} \, L^2 \, T_{ab}, \, = \, 
\Big[\frac{6}{C} \, L^2 \, \frac{1}{\hbar c} \, T_{ab}\Big],
\end{eqnarray}
with the constant $\Lambda$, we see, actually
playing the role of the cosmological constant.

Now,
since Einstein's field equations read
$
G_{ab} + \Lambda g_{ab} = 8 \pi \, T_{ab} 
\Big[ = 8 \pi \, l_p^2 \, \frac{1}{\hbar c} \, T_{ab}\Big],
$
(\ref{nullq41.6}) is actually Einstein's equations
if
$
l_p^2 = L^2/\big({\frac{4}{3} \pi \, C}\big).
$
Using this in the extremality condition (\ref{nullq41.4})
we get

\begin{eqnarray}\label{nullq42.7}
-\frac{1}{6} C L^2 \, 8\pi l_p^2 \,
\frac{R_{ab} l^a l^b}{8\pi l_p^2} \, + \,
L^4 \, T_{ab} l^a l^b &=& 0
\nonumber \\ 
L^4 \, \, \bigg(-\frac{R_{ab} l^a l^b}{8\pi l_p^2}\bigg) \, + \, 
L^4 \, T_{ab} l^a l^b &=& 0
\nonumber \\
H_G + H_m &=& 0
\end{eqnarray}
\cite{Pad06, Pad10, Pad12, Pad16},
with
$
H_G \equiv - R_{ab} \, l^a l^b/(8\pi l_p^2) =
\Big[ - \frac{1}{8 \pi} \, \frac{\hbar c}{l_p^2} \, R_{ab} \, l^a l^b\Big]
$
and
$
H_m \equiv T_{ab} \, l^a l^b.
$

Coming back to the specifications just below
equation (\ref{nullq44.3}) (with $k^a = l^a$),
we see that 
$
H_m = \delta Q_H/(A_H \, l),
$
is the heat flux brought by matter
when crossed by the horizon in time $dt = l$.
Indeed, 
$H_m$ is energy (in matter's frame)
supplied by matter
per unit (engulfed) proper volume,
or also 
matter energy per unit 
cross-section area of horizon  
per unit (matter) crossing time.

The $H_G$ term can also be
a heat flux, but relative to spacetime instead of matter.
It can be interpreted as
the heat generated by spacetime-induced viscous stresses
on our congruence considered as a null fluid, 
in reaction to the presence of matter \cite{PadO, KolekarPaddy}.
This can be understood as follows.
We start from the identity (see e.g. \cite{PadZ}, equation (A.56) there)

\begin{eqnarray}\label{nullq48.1}
\nabla_a(\theta \, l^a) + \nabla_a l^b \, \nabla_b l^a - \theta^2 =
- R_{ab} \, l^a l^b,
\end{eqnarray}
valid for a generic null congruence,
which we apply to our affine congruence
(in general non-affine parameterization 
$\theta = \nabla_a l^a + \kappa$, with $\kappa$
defined by $l^b \nabla_b l^a = \kappa \, l^a$;
in our affine parameterization $\theta = \nabla_a l^a$).
From
$
\nabla_a l^b \, \nabla_b l^a = \sigma_{ab} \sigma^{ab} + \frac{1}{2} \theta^2
$
in our case 
(as can be easily worked out, starting e.g. from
\cite{Poisson}), 
equation (\ref{nullq48.1}) becomes

\begin{eqnarray}\label{nullq48.2}
\nabla_a(\theta \, l^a) + \sigma_{ab} \sigma^{ab} - \frac{1}{2} \theta^2 =
- R_{ab} \, l^a l^b,
\end{eqnarray}
or

\begin{eqnarray}\label{nullq48.7}
\nabla_a(\theta \, l^a) + \sigma_{ab} \sigma^{ab} - \frac{1}{2} \theta^2 = 0
\end{eqnarray}
in vacuum.

On the other hand,
the heat flux $H$ generated by viscous stresses
in a fluid can be expressed as \cite{LandauLifshitzFM, MTWPrinceton}

\begin{eqnarray}\label{viscous heat flux}
H = 2 \, \eta \, \sigma_{ab} \sigma^{ab} + \zeta \, \theta^2,
\end{eqnarray}
with $\eta$ and $\zeta$ coefficients of shear and bulk viscosity
respectively.
At equilibrium, 
one expects no entropy generation by viscous heating,
and thus, in case the viscous stresses are the only source
of entropy,

\begin{eqnarray}\label{nullq52.2}
2 \, \eta \, \sigma_{ab} \sigma^{ab} + \zeta \, \theta^2 = 0.
\end{eqnarray}

Now,
from general results concerning null fluids and black hole membrane paradigm
\cite{Damour1, Damour2, MembraneP}
one has 
\begin{eqnarray}\label{eta zeta}
\eta = \frac{1}{16 \pi} \, , 
\, \, \, \zeta = - \frac{1}{16 \pi}.
\end{eqnarray}
This implies that
at equilibrium 
we must have
$
\sigma_{ab} \sigma^{ab} = \frac{1}{2} \, \theta^2.
$
Using this in (\ref{nullq48.7})
we see that the equilibrium condition corresponds to
$
\nabla_a (\theta \, l^a) = 0.
$
For null hypersurfaces generic
we have in general no reason to
expect $\nabla_a (\theta \, l^a) = 0$,
but based on the above it seems we can say that 
a null hypersurface is actually 
a description of a null fluid at equilibrium when 
$\nabla_a (\theta \, l^a) = 0$. 

The introduction of matter brings additional stresses in the null fluid
as can be inferred from looking at (\ref{nullq48.2}).
We have a new source of heating and then of entropy.
These are stresses induced by spacetime (from the presence of matter)
and the heat flow they induce can be read in (\ref{nullq48.2})
(using (\ref{viscous heat flux}) and (\ref{eta zeta})) 
as $\frac{R_{ab} \, l^a l^b}{8 \pi}$.
A new equilibrium is reached when the total entropy generation,
including that sourced by spacetime-induced heat,
vanishes.
This corresponds to 
$
2 \, \sigma_{ab} \sigma^{ab} -\theta^2 + 2 \, R_{ab} l^a l^b~=~0 
$
and
again 
$
\nabla_a (\theta \, l^a) = 0.
$
We thus have equilibrium when

\begin{eqnarray}\label{nullq53.1}
- R_{ab} \, l^a l^b \, = \, \sigma_{ab} \sigma^{ab} - \frac{1}{2} \, \theta^2,
\end{eqnarray}
and then when the viscous heating in presence of matter, 
which is 
$
H_G =
2 \, \eta \, \sigma_{ab} \sigma^{ab} + \zeta \, \theta^2
= \frac{1}{8 \pi} \, \sigma_{ab} \sigma^{ab} - \frac{1}{16 \pi} \, \theta^2
\ne 0,
$
is given by

\begin{eqnarray}\label{nullq53.2}
2 \, \eta \, \sigma_{ab} \sigma^{ab} + \zeta \theta^2 
\, = \,
-\frac{R_{ab} l^a l^b}{8 \pi}
\, \Bigg[ = \, 
- \frac{\hbar c}{l_p^2} \, \frac{R_{ab} l^a l^b}{8 \pi}\Bigg]
\end{eqnarray}
\cite{PadO, KolekarPaddy, PaddySumanta}.

Given the meaning of $H_G$ and of $H_m$
the extremality condition (\ref{nullq42.7})
acquires an interpretation as total heat flux in the null 
hypersurface = 0 \cite{Pad06, Pad10, Pad12, Pad16}.
That is: the heat flux brought in by matter crossing the null hypersurface,
call the latter a horizon,
plus the heat flux associated to the internal stresses of the horizon
must vanish, 
a result which is known as `dissipation without dissipation'
or also as `zero-dissipation principle' \cite{Pad08, Pad10,Pad12, Pad16}. 
But,
we showed that 
(provided we set a suitable link between the constant $C$
and $l_p$)
the extremality condition means
Einstein's equations (with cosmological constant).
The statistical extremization procedure we have described 
then shows that what Einstein's equations really are is 
the statement that total heat flux is 0 \cite{Pad06, Pad10, Pad12, Pad16}.

\subsection{On the nature of Einstein's equations}

The importance
of this result, by Paddy, can hardly be overstated.
This way to look at Einstein's equations indeed results in
something far more satisfactory than the conventional one.
Indeed the equality 
$G_{ab} = 8\pi \, T_{ab}$ (ignoring here the cosmological term)
is between two quantities
which could not be more different in their own nature:
on one side exact pure geometry, on the other all the stuff related
to matter; we might call it an equality between marble (geometry) 
and wood (matter quantum fields) following Kaku \cite{Kaku}, 
or between apples and oranges in Paddy's own words \cite{Pad20},
and this without any link to all the thermodynamic meanings
found for gravity in the many years from when
these equations were first formulated. 

Precisely this unsatisfactory element of the conventional perspective
lead Paddy to look for a description of gravity
embodying thermodynamics since start,
seeking for a picture of field equations
as what comes from the extremization of a suitable
entropy functional capable to include matter and gravity.
Well before the minimum-length spacetime entered the scene, 
he and collaborators found this functional,
and the extremization turned out to consist 
in imposing an equality between gravitational and matter 
entropy variations on null hypersurfaces,
and resulted equivalent to field equations
(with cosmological constant) \cite{PadF, PadG_} (see also \cite{PadN}). 

With the advent of the minimum-length description of spacetime
this functional nicely
was found to coincide, up to a total divergence  
which can be ignored in the extremization procedure,
with $N_G + N_m$, i.e. the dofs of spacetime and matter 
described above \cite{Pad20}.
Thus,
even in the absence of a full theory of quantum gravity
and then of an account of gravity intrinsic dofs,
the approach with minimum-length spacetime is capable of giving
to that thermodynamic description of field equations
a statistical-mechanical basis,
sort of a hint towards 
a possible statistical origin of gravity
along what sought for from entropy bounds
(cf. the paragraph just above that of equation (\ref{15.2})).  

Looking at equation (\ref{nullq44.3}) describing 
Jacobson's result \cite{Jacobson},
we recognize in it the terms $H_m$ and $H_G$ (with $k^a = l^a$),
and that equation too can actually be read
as a total heat flux = 0. 
%
%
But, 
building on Jacobson's,
Paddy's result opens new avenues of research by
going definitely beyond it in at least two respects.

One is the fact that in Jacobson's result \cite{Jacobson}
the heat flux $H_m$ is balanced by the heat flux associated
to entropy variation of the horizon {\it assuming} this entropy
is proportional to horizon's area.
In Paddy's derivation on the contrary, 
there is no such assumption:
the heat flux $H_G$ is that computable 
from the viscous stresses of the 
null surface making up the horizon,
in reaction to the presence of matter.

The other one
is the fact that Jacobson's is a thermodynamic balance equation;
Paddy's is instead a statistical-mechanical one.  
As such it arises
from a description in terms of micro dofs.
In this respect Paddy's result goes beyond Jacobson's
in the same sense
that statistical mechanics goes beyond thermodynamics.
It provides indeed a more accurate picture of the physical world,
in that it takes note, and then uses in the mathematical description,  
kind of particle nature of the constituents. 
With Paddy we can talk of particle-like constituents for spacetime,
meaning with this what is responsible for its micro dofs.

\subsection{Classical gravity can not be considered to be
the $L\to 0$ ($\hbar\to 0$) limit} 

The derivation of field equations is universally done
from extremization of a suitable action
with respect to the field,
and
gravity is no exception with the metric being the field.
One would like to know how to reconnect
an extremization of this kind
with finding the field equations
as an extremization of entropy 
along the lines described above.

In case of Einstein's equations the Lagrangian in the action
is the Ricci scalar $R$. 
Considering things
at thermodynamic level 
(as opposed to a statistical-mechanical one),
the relation between entropy extremization 
and derivation from variation of the action
has been discussed in \cite{TFF}.
What has been found is that
the thermodynamic extremization can 
be seen as mathematically equivalent to varying 
a modified, suitable
action (the `Augmented variational principle' of 
\cite{FatibeneFerrarisFrancaviglia})  
with respect to the field.

One would however also investigate
which kind of connection exists 
at the statistical-mechanical
level, meaning with this what we obtained
with the minimum-length description of spacetime.
Crucial for this is to find out
the expression $R_{(q)}$ of the 
Ricci scalar in the qmetric.

The calculation of $R_{(q))}$ is not an easy task
since the qmetric covariant derivative
has terms additional to ordinary covariant 
derivative (see right below equation (\ref{affineq}))
and with further
manipulations the algebraic complications diverge.
Fortunately,
a relatively quick way to compute it has been
devised.
The basic idea is
to avoid the calculation of the components
of the Ricci tensor (going then to take the trace)
and to calculate instead directly
the Ricci scalar resorting
to Gauss-Codazzi relations for the equigeodesic hypersurfaces  
\cite{Pad01, JaffinoKot, KotG}.

The nonlocal Ricci scalar $R_{(q)} = R_{(q)}(p, P)$ 
one thus obtains has the property that
if we take 
the coincidence limit $p \to P$ 
and then consider the limit $L \to 0$
we get (in D-dim spacetime)

\begin{eqnarray}\label{Rq t/s}
R_{(q)}(p, P) \to \epsilon \, D \, R_{ab} t^a t^b 
\end{eqnarray} 
for spacelike/timelike geodesics with 
(unit) tangent $t^a$ \cite{Pad01, JaffinoKot}.
This has been extended 
to include the case of null geodesics \cite{PesP}
(from Gauss-Codazzi as applied to null
hypersurfaces \cite{Gem, CP})
getting

\begin{eqnarray}\label{Rq null}
R_{(q)}(p, P) \to \, (D-1) \, R_{ab} l^a l^b,
\end{eqnarray}
with $l^a$ the null tangent vector.

A most intriguing feature of these results
is that the coincidence-limit
qmetric Ricci scalar does not tend to $R$
when $L\to 0$
as one would have instead naively expected.
We do not recover ordinary spacetime
when letting $L\to 0$. 
If we think of the minimal length as 
$L = C \, l_p$ with $C$ a constant,
this means
that we do not get classical spacetime 
in the $\hbar \to 0$ limit.
Even more so
in the general situation in which $L$ may not
vanish with $\hbar$.

Moreover, the limiting values are proportional to terms like
$R_{ab} t^a t^b$ and $R_{ab} l^a l^b$,
namely right the terms we get
in equation (\ref{expansion density G}) above
when counting the micro dofs of
gravity.
Things go as if endowing spacetime with a minimum length,
read giving spacetime quantum characteristics,
turns the Lagrangian into counting gravitational dofs,
and the extremization through a variational principle
into an extremization of entropy. 
For any nonvanishing $L$ this statistical-mechanical machinery works,
providing 
the field equations of classical gravity.
From 
inspection of equation (\ref{nullq41.6}),
note
that 
the field equations become singular
in the $\hbar \to 0$ limit.
This shows that we can not get classical gravity
when $\hbar \to 0$;
to have classical gravity (the field equations (\ref{nullq41.6}))
we definitely need $\hbar \ne 0$ (cf. \cite{Pad06}).  

On top of this,
since this correspondence 
\{action extremization\} $\to$ \{entropy extremization\}
remains there no matter
how small $L$ is,
the terms $R_{ab} t^a t^b$ and $R_{ab} l^a l^b$,
with all their thermodynamic significance,
are sort of echo and witness
of the underlying quantum structure \cite{Pad01},
with the latter made this way visible even when $L$ 
(read the Planck length)
is hopelessly (for direct experimental detection) small.

\subsection{Clues about a small-scale quantum description}

We see in the formulas above that 
the $L\to 0$ value of the qmetric Ricci scalar
depends on the direction of approach to $P$.
Since the Ricci scalar ought to be considered 
as a quantity given with the manifold and $P$,
this of the dependence on the direction 
is a peculiar feature that calls 
for further understanding 
(a proposed interpretation is in \cite{PesQ}).
The ordinary Ricci scalar $R$ is recovered (in the limits above)
when consistently averaging over all geodesics 
from $P$ \cite{PesR}.
All this might hint
to that what the minimum-length metric (to some extent) captures
is a quantum structure for spacetime at $P$,
and that 
it is by averaging,
or taking expectation values we would say,
that we can reconnect
$R_{(q)}$ with the ordinary $R$.
This structure would be encoded and witnessed in the terms 
$R_{ab} t^a t^b$ and $R_{ab} l^a l^b$
and their thermodynamic baggage.
One can also try to characterize this local quantum structure
in an operational way, 
thinking of the event $P$ as a coincidence between
a quantum reference body ($B$)  and a test particle ($T$).
In this case,
pure states for the system $B {\otimes} T$
(corresponding to definite directions of approach to $P$)
turn out to be mixed for system $B$ alone
(due to the nonvanishing limiting transverse area
which gives an irreducible finite probability for $B$ to wrongly
guess the arrival direction of $T$).
Considering photons as $T$
the quantity $R_{ab} l^a l^b$ emerges once more,
this time in the average information gain $I$
associated to $B$ finding the photon on nominal geodesics
(tangent $l^a$) when measuring along $l^a$;
one finds $I \propto L^2 \, R_{ab} l^a l^b$ \cite{PesU}.      

These quantum features would stay there
no matter the smallness of $L$ (read $\hbar$ if we think
of the Planck length $l_p$ for $L$),
this reminding of Bell inequalities,
whose violation certifies nonlocality,
but with the size of the violation being unrelated
to the value of the Planck constant.
This offers in principle a hope to test quantum features 
of gravity also in circumstances in which any effect
proportional to $l_p$, or its powers, is hopelessly small
as is surely the case in the lab,
thing that
intriguingly resonates with recent new proposals
to check the non-classicality of gravity in the lab
\cite{BosA, MarlettoVedral, CR, BosB}.
 
As for 
possible signatures of a minimum-length metric,
we have to consider that these
come with the fact that
the latter is a way 
to coherently embody the metric description
of spacetime with a limit length, plausibly of quantum origin, 
in the small scale.
Then any circumstances in which
quantum effects are expected to induce 
deviations from the classical metric
are suitable test-beds for the minimum-length metric,
and the use of the latter in place of ordinary metric
should affect, and hopefully refine
the theoretical predictions.
(Very early) cosmology and extreme astrophysical events 
might be thought then
as the main topics where to find signatures or make convenient use  
of a modification of the ordinary metric to a minimum-length metric.

In particular, on the basis of the back-on-the-envelope arguments 
mentioned above concerning singularity-formation avoidance
(the two paragraphs right after Eq. (\ref{thetaq_0})),
one can expect 
that in the minimum-length metric 
the prototypical
Oppenheimer-Snyder black-hole collapse brings to shrink matter 
to a finite limit area and 
not to form a singularity
(this is what expected on the basis of those arguments, 
but might deserve a full-fledged proof).
Evaporation will then shrink the horizon
and the process might be expected to effectively halt when the horizon area
becomes small enough to be comparable with the limit area;  
this might likely correspond to a black hole remnant
and the minimum-length metric might be a tool
for the description of its metric.
This could be possibly helpful when coming e.g. to study 
dark matter along the well-known 
idea that it might be explained
in terms Planck-size black-hole remnants
(problematic idea in principle \cite{Susskind_BHremnants, Hawking_BHremnants}, 
but possibly viable, according to successive accounts 
\cite{Calmet, White holes, RovelliVidotto}).

On the same grounds,
following back in time the evolution of a LFRW universe
as traced by worldlines of particles comoving with the universe,
if we use the minimum-length metric
we may expect to reach a limit configuration
corresponding to the limit `area' (spatial volume actually)  
(\ref{finite area s/t}).
As in the collapse case described above,
this would hint to circumstances in which
the universe can not effectively shrink (back in time) 
any longer, and the minimum-length metric
might be useful in describing its metric properties
at these and (not too much) later times
when typical distances between inhomogeneities 
would be strongly affected by the deviation from classicality
of the metric.
This would correspond to the first stages of 
the very early universe, right after the Planck epoch,
if the limit length $L$ is the Planck length $l_p$;
but it could be relevant for much later stages 
(still in the very-early-universe epoch) if 
$L$ is significantly larger than $l_p$.
Detailed calculations with the minimum-length metric 
might be worth pursuing within the LFRW models.

\section{Conclusions}

What we have done in the paper
has been to go through a series of results
connecting gravity and thermodynamics/statistics/quantum information
starting from 
the fundamental result \cite{Jacobson}
by Jacobson.
The path chosen reflects the personal line of research of the author
entangled in multiple ways with (small) part of own research by Paddy
in the last decade.

In the first part,
specifically of thermodynamic flavor, 
the focus has been on that 
even if we try to keep nonlocality away
as long we can in the consideration of gravity,
resorting in particular to
matter in the hydrodynamic approximation
(thus with energy and entropy microscopically related), 
we end up unavoidably with a nonlocal picture,
thus kind of a nonlocal spacetime, 
at least at the smallest scales (the Planck length).
The stress is on that gravity exhibits
an irreducibly nonlocal character.
This has been discussed building basically
on entropy bounds,
with null hypersurfaces and photon gases playing a major role. 

The second part has been devoted to describe
the explicit implementation of this nonlocality 
in the description of spacetime,
with the so called minimum-length metric or qmetric
\cite{KotE, Pad01, JaffinoKot}.
We have seen how this is accomplished 
through use of bitensors, thus objects depending on two separated events,
in particular through a metric bitensor $q_{ab}$
which gives that in the modified (or enriched with a microstructure)
spacetime there is a lower limit $L$ 
to the separation distance between two events in their coincidence limit.
Several interesting things do happen in the qmetric spacetime.
Among them one stands out (and has been given specific attention
in the paper): The existence of nonvanishing limit areas
attached to every event in spacetime.
This is what
brings to introduce micro dofs for gravity
and enables a statistical-mechanical description of 
matter+gravity system
with a max-statistical entropy account of field equations;
this has been spelled out in some detail in the paper
for the case of null separated events.
Very intriguingly, this statistical description keeps staying
there also in the $L\to 0$ limit.  

The gross picture we can get from this
is that
the field equations of gravity
might be in essence a statistical-mechanical result,
and as such ought to
be expressible completely using statistical-mechanical, not geometric, 
concepts.
Spacetime and the geometric description of gravity we are used to, 
emerge
when circumstances are such that what we observe is due 
to cumulative effects over many micro dofs 
and a continuous description
becomes viable, what we call the thermodynamic limit
(cf. \cite{ChaP} and references therein).
Key to this picture would be a clear understanding of the 
emergence of time; 
we believe that something so much intriguing/promising regarding this
and possibly work on 
is Paddy's recent work \cite{Paddy}.

On top of this,
gravity appears to be inherently quantum
(as hinted to by the
persistence of the statistical-mechanical construct in the
$L\to 0$ limit; the `classical description' does not come up
in the $\hbar \to 0$ limit:
it is there for (any) $\hbar \ne 0$ but we do not have it any longer 
(the field equations become singular)
in the $\hbar\to 0$ limit).
And even though nonlocality unavoidably appears
(in whichever circumstances) in the small scale limit,
this does not preclude it to come up at more mundane or, also, lab scales
(as mentioned at the end of previous section) if suitable
circumstances can be considered.
``Gravity is quantum mechanical at all scales'' \cite{Pad06} after all,
using again Paddy's words, enlightening as ever.

{\it Acknowledgments.}
I thank  Francesco Anselmo for 
drawing attention to one of the references.
This work was supported in part by INFN grant FLaG. 

{\it Data availability statement.}
All data generated or analysed during this study are included in this 
published article.


\end{document}